\documentclass{aa}
\usepackage{graphicx}
\usepackage[colorlinks=true, linkcolor=blue, citecolor=blue, urlcolor=blue]{hyperref}

\usepackage{txfonts}
\usepackage{array}
\usepackage{booktabs}
\usepackage{float}
\usepackage{placeins}
\usepackage{enumitem}
\usepackage{pgfplots}
\pgfplotsset{compat=1.18}

\usepackage{natbib}
\hypersetup{
    colorlinks=true,
    linkcolor=blue,
    citecolor=blue,
    urlcolor=blue
}
\usepackage{xcolor}
\begin{document}

   \title{Circumgalactic medium depletion drives satellite quenching in IllustrisTNG}

   \titlerunning{CGM depletion drives satellite quenching in IllustrisTNG}

   \author{Natan de Isídio
        \inst{1}\fnmsep\thanks{\href{www.natanisidio.com}{Natan.Isidio@eso.org}}
          \and Paola Popesso \inst{1}
          \and Sandro Tacchella\inst{2,3}
          \and Anna Pasquali\inst{4}
          \and Ilaria Marini\inst{1,5}
          \and Daudi Mazengo \inst{1,6}
          \and \\Victoria Toptun \inst{1}
          \and Sean McGee\inst{7}
}

   \institute{European Southern Observatory, Karl-Schwarzschild-straße 2, 85748 Garching bei München, Germany
    \and Kavli Institute for Cosmology, University of Cambridge, Madingley Road, Cambridge CB3 0HA, UK
    \and Cavendish Laboratory, University of Cambridge, 19 JJ Thomson Avenue, Cambridge, CB3 0HE, UK
    \and Astronomisches Rechen Institut, Zentrum f\"ur Astronomie der Universit\"at Heidelberg, M\"onchhof-straße 12-14, D-69120 Heidelberg, Germany
    \and Fakultät für Physik, Ludwig-Maximilians-Universität München, Scheiner-straße 1, 81679 München, Germany
    \and College of Natural and Mathematical Sciences, University of Dodoma, Benjamin Mkapa Road, 41218 Iyumbu, Dodoma, Tanzania
    \and School of Physics and Astronomy, University of Birmingham, Birmingham B15 2TT, UK
    }

   \date{Received August 2026; accepted --}
 
\abstract{
Satellite galaxies dominate the quenched population at low stellar masses ($M_\star $$\,\lesssim\,$$ 10^{10}\,\rm M_\odot$), yet identifying which processes shut down their star formation, their relative importance, and on what timescales, remains a central problem in galaxy evolution. We use MaNGA-like mock galaxies from IllustrisTNG to dissect different satellite quenching pathways, paying special attention to the role of the circumgalactic medium (CGM) during quenching phase. 
With a merger-tree-based algorithm, we reconstruct the baryonic, dark matter, structural, and chemical histories of $\sim$7\,300 spatially resolved galaxies (2\,800 satellites) along their \textsc{SubLink} trees, using time since infall as the physical axis along which quenching unfolds.
Satellites retain regular, rotation-supported stellar kinematics throughout quenching, with disturbed velocity fields confined to systems with $M_\star $$\,\lesssim\,$$ 10^{10.5}\,\rm M_\odot$. 
For the first time, we present the coupled time evolution of the depletion of both the hot and cool gas reservoirs after infall: satellites lose $\sim$90\% of their hot CGM within $\sim$$4.2^{+0.6}_{-0.6}$\,Gyr, increasing with residence time and independent of stellar mass. 
The hot gas mass correlates strongly with SFR, establishing the CGM as the long-term fuel reservoir, unlike quenched centrals, which retain massive hot halos likely maintained by AGN feedback. Present-day quenched satellites were accreted earlier than star-forming ones (6.5$^{+0.3}_{-0.3}$ vs.\ 4.3$^{+0.3}_{-0.3}$\,Gyr ago), forming stars for at least $\sim$3\,Gyr after infall before declining sharply, consistent with a delayed-then-rapid quenching scenario. Losing little stellar mass, yet with their gas depleted and their dark matter and metal-poor stellar outskirts tidally stripped, satellites emerge more compact and metal-rich than centrals at fixed mass. Our results suggest the gradual erosion of the hot CGM as the key link connecting infall to the slow shutdown of star formation.}

\maketitle

\section{Introduction}
\label{Sec1}

Understanding why galaxies stop forming stars, a process known as quenching, remains one of the central unsolved problems in galaxy evolution. The observed growth of the quenched population since at least $z$$\,\sim\,$$2$, together with the sharp bimodality in color, star formation rate (SFR), morphology, and kinematics, implies that star formation is not simply exhausted stochastically, but is regulated, and ultimately shut down, by a limited set of physical pathways acting across a broad range of halo masses and environments \cite[e.g.,][]{Bell+04, vandenBosch+08,Peng+10, Muzzin+14, Jaffe+15, Wetzel+15, Popesso+15, Tacchella+15,Tacchella+16,Oman+16, Bluck+20, Trussler+20, Cortese+21, Tacchella+22a, deIsidio+24,Baxter+25, deIsidio+26}. 
The difficulty, though, is not in naming candidate mechanisms, but in disentangling their relative roles, timescales, and observable imprints \cite[e.g.,][]{deIsidio+26}.

Preventing fresh gas accretion (hereafter referred to as starvation), removing gas already bound to galaxies (via ram-pressure stripping, turbulent viscous stripping, or tidal interactions), stabilizing or heating halo gas through active galactic nucleus (AGN) feedback (quenching maintenance), and dynamically reshaping galaxies via mergers and structural compaction can all move systems off the star-forming main sequence. 
However, these channels do not necessarily operate independently; most likely, the dominant pathway may involve a sequence of processes rather than a single event \cite[e.g.,][]{Larson+80,Cortese+19,Bluck+20,Donnari+21,Visser-Zadvornyi+25,deIsidio+26}.

A useful approach is to separate the problem by the galaxy's role within its host dark-matter halo. For central galaxies, quenching is widely linked to the emergence of a stable hot halo near the peak of baryon conversion efficiency, often reinforced by black hole feedback that suppresses late-time cooling and prevents rejuvenation \cite[e.g.,][]{Croton+06,Bower+06,Weinberger+17,Pillepich+18,Bluck+20,Xie+20}. In this regime, quenching is best understood as a maintenance problem: galaxies do not simply consume their cold gas, but are also prevented from cooling the hot gas reservoir.

For satellites, the physical picture is inherently more complex. 
In the local Universe, satellites dominate the quenched population below $M_{\star}$$\,\sim\,$$10^{10.5}$\,M$_{\odot}$ and are substantially more likely to lie below the star-forming main sequence than centrals of comparable mass \citep[e.g.,][]{Popesso+19a,Cortese+21,Popesso+23}. 
For instance, infall into a larger halo may cut off cosmological gas supply, expose galaxies to the thermodynamic pressure of the host circumgalactic (CGM) or intragroup/cluster medium (IGrM/ICM), and enhance the role of tidal perturbations and satellite--satellite encounters. For this reason, observations and simulations indicate that satellite quenching cannot be reduced to a single mechanism: low-mass satellites are particularly sensitive to environmental effects, often showing signatures consistent with outside-in suppression, while more massive satellites can retain substantial similarities to quenched centrals and may require the combined action of environmental processing and internal feedback \cite[e.g.,][]{Wetzel+13,Peng+15,Pasquali+19,Bluck+20,Baxter+25}. 

In this context, the hot circumgalactic reservoir of galaxies is not a side issue but a central one.
Affecting the CGM, either by stripping the hot gas reservoir of a satellite upon infall (starvation), or by halo heating due to AGN in centrals (quenching maintenance), is fundamentally about regulating the exchange between galaxies and their gaseous surroundings \cite[e.g.,][]{McCarthy+08,Tumlinson+17,Peroux+20,Cortese+21}. 
Simulations increasingly support this view. In TNG-like and FLAMINGO-like models, quenching is closely linked to changes in the inner CGM and to the onset of effective black-hole feedback \cite[e.g.,][]{Joshi+21,Donnari+21,Donnari+21b,Rohr+24,Lim+25}.
For satellites, infall into a hot host halo changes not only the rate of gas accretion but also the physical accessibility of the remaining gas reservoir, making the time since infall a physically motivated axis along which to interpret quenching observables \cite[e.g.,][]{Pasquali+19}. This is especially relevant for the mass--metallicity evolution of satellites, because metallicity enhancement, stellar mass loss, and hot-gas depletion need not occur synchronously: gas removal may be fast, the stellar response delayed, and chemical signatures cumulative.

A picture is therefore emerging in which stripping and starvation are not competing alternatives but successive stages of a single quenching sequence, distinguished primarily by the timescale and the gas phase on which they act. Shortly after infall, environmental forces act most efficiently on the least gravitationally bound material (the hot CGM and the extended, diffuse outskirts of the disk) which can be removed relatively rapidly, often within $\sim$1--2\,Gyr and close to first pericentric passage \cite[e.g.,][]{Gunn+72,Wetzel+13,Rohr+24}.
This early stripping (ram pressure and tidal) does not by itself terminate star formation: a satellite can continue forming stars for several Gyr from the cold gas already within its disk, even as the surrounding reservoir is eroded. 
In this quenching scenario, what ultimately quenches the galaxy is the loss of the hot halo that would otherwise cool and replenish that disk, so that quenching proceeds as a delayed, gradual exhaustion of an increasingly closed reservoir rather than an abrupt shutdown \cite[e.g.,][]{vandenBosch+08,Wetzel+13,Peng+15,Cortese+21}. 
Such a sequence is expected to leave a distinctive set of imprints that can be tested jointly: gas reservoirs that decline sharply after infall while the stellar mass remains essentially frozen, stellar metallicities that rise as low-metallicity accretion is cut off and metal-poor outer material is preferentially stripped, an accompanying structural compaction as the loosely bound outer layers are tidally stripped, and a central stellar body whose ordered kinematics survive relatively intact, since ram-pressure gas removal exerts no force on the collisionless stellar component while tides preferentially strip the weakly bound outskirts and leave the rotation-supported core largely undisturbed \cite[e.g.,][]{Pasquali+12,Schaefer+17,deIsidio+26}. Disentangling this sequence requires following all of these observables together as a function of time since infall, which is what applying a straightforward approach to a forward-modeled cosmological hydrodynamical simulation makes possible.

In this paper, we use MaNGA-like mock galaxies from the IllustrisTNG simulation to dissect the relative importance of different quenching pathways, paying special attention to the role of the CGM during the quenching phase.
In Section~\ref{Sec2}, we describe our mock sample and the algorithm developed to analyze the environment and merger history of all galaxies in our sample. 
Section~\ref{Sec3} outlines the methodology used to identify and quantify kinematic asymmetries, as well as a brief discussion on the gas kinematics in TNG50. 
In Section~\ref{results}, we present the main results, giving particular attention to the role of the CGM in shutting down star formation.
Finally, in Section~\ref{Sec5} we discuss each of the results presented in Section~\ref{results} along with our understanding of their implications on galaxy quenching.
By combining chemical evolution, gas accretion/loss history, infall time, and kinematic disturbances within a single forward-modeled framework, the main aim of this work is to provide evidence of what are the most important mechanisms behind satellite quenching.

Throughout this work, we adopt $Z_\odot$$\,=\,$$0.0142$ \cite[][]{Asplund+09}, and a flat cosmology based on the standard $\Lambda$CDM model is employed, with cosmological constraints set to $H_0$\,=~(67.8~$\pm$\,0.9) km~s$^{-1}$\,Mpc$^{-1}$, $\Omega_m=0.308$$\,\pm\,$$0.012$, $\Omega_{\Lambda}=0.691$$\,\pm\,$$0.012$, and a cosmological baryon fraction of $f_b~=~\Omega_b/\Omega_m = 0.1573$ \cite[][]{PlanckCollaboration+16}.

\section{Data and sample}
\label{Sec2}
\subsection{TNG50 simulation}
The TNG50-1 simulation is part of the IllustrisTNG project, a suite of cosmological magnetohydrodynamical simulations designed to model galaxy formation and evolution \cite[][]{Pillepich+19,Nelson+19}. 
TNG50-1 is the highest-resolution run within the IllustrisTNG suite, featuring a cubic volume with a comoving side length of 51.7~Mpc, incorporating updated models for AGN feedback and galactic winds \cite[][]{Weinberger+17,Pillepich+18}.
It contains 2160$^3$ dark matter and baryonic particles, with mass resolutions of approximately $4.5 \times 10^5~\rm{M}_\odot$ for dark matter and $8.5 \times 10^4~\rm{M}_\odot$ for baryons. The gravitational softening length reaches 72~pc at $z$ = 0, enabling detailed investigations of small-scale galactic structures and internal dynamics \cite[][]{Pillepich+19, Nelson+19}.
The simulation follows a $\Lambda\rm{CDM}$ cosmology consistent with \cite{PlanckCollaboration+16} results, adopting $\Omega_{m}$ = 0.3089, $\Omega_{\Lambda}$ = 0.6911, $\Omega_{b}$ = 0.0486, $h$ = 0.6774, and $\sigma_8$ = 0.8159. 
It begins at $z$ = 127 and evolves to the present day ($z$ = 0), including key astrophysical processes such as primordial and metal-line cooling, heating by the extragalactic UV background, stochastic star formation, and feedback from supernovae, AGB stars, and supermassive black holes \cite[][]{Weinberger+17,Pillepich+18}.
Haloes within TNG50-1 are identified using the Friends-of-Friends (FoF) algorithm \cite[][]{Davis+85}, while subhaloes are recognized as gravitationally bound structures via the SUBFIND algorithm \cite[][]{Springel+01, Dolag+09}. 
Merger histories of subhaloes are traced across snapshots using the \textsc{Sublink} merger trees \cite[][]{Rodriguez-Gomez+15}. The simulation's resolution and physical modelling make it particularly well-suited for analyzing satellite galaxies within Milky Way-mass haloes and investigating galaxy evolution at sub-kiloparsec scales over cosmic time.

\subsection{Mock MaNGA sample: MaNGIA}
To directly compare theoretical predictions with previous works based purely on observations, we use the MaNGIA sample \citep[Mapping Nearby Galaxies with IllustrisTNG Astrophysics;][]{Sarmiento+23}. The dataset consists of over 10\,000 mock integral field spectroscopic (IFS) datacubes designed to emulate the observations of the MaNGA survey \citep{MaNGA+15,MaNGA+22}. 
MaNGIA forward-models TNG50 galaxies through the MaNGA instrumental pipeline, with its sample carefully matched to the mass, size, and redshift distribution of the MaNGA Primary, Secondary, and Color-Enhanced samples to minimize selection biases.

The MaNGIA mock datacubes are constructed by assigning MaStar SSP spectra to TNG50 stellar particles, incorporating dust attenuation and instrumental effects (seeing, fibre dithering, noise), and are processed with \texttt{pyPipe3D} to ensure consistency with the original MaNGA data reduction pipeline.
The final sample successfully reproduces the global scaling relations observed in MaNGA, such as the tendency for more massive galaxies to be older and more metal-rich, as well as the expected increase in velocity dispersion with stellar mass \citep[see Fig.\,10 of][for a comparison of the main galaxies' properties in MaNGA and MaNGIA]{Sarmiento+23}.
Despite broad agreement, MaNGIA exhibits some important discrepancies: low-to-intermediate mass galaxies (M$_\star $$\,\lesssim\,$$10^{10}\,$M$_\odot$) are older and more metal-rich than observed, while the most massive galaxies tend to show lower velocity dispersions (>250\,$\mathrm{km\,s^{-1}}$) and show an overabundance of fast rotators, suggesting TNG50 produces excessive massive disks compared to the observed MaNGA sample.
Readers are referred to \cite{Sarmiento+23} for a detailed overview on MaNGIA-mock catalog.

Rather than adopting the full MaNGIA catalog, we apply extra selection criteria to obtain a clean, non-redundant sample of galaxies with reliable kinematics. 
By construction, MaNGIA repeats galaxies: since TNG50 does not contain enough objects to reproduce the full MaNGA mass--size--redshift distribution, individual subhalos are observed from up to six viewing directions and included multiple times. 
To avoid counting the same physical system more than once, we keep only datacubes with $\mathtt{view}=1$, i.e. galaxies observed along the positive $x$-axis of the simulation box. Since this axis is arbitrary with respect to each galaxy's intrinsic orientation, it provides an unbiased line of sight while ensuring that every subhalo enters our sample exactly once. 
In addition, we require each object to contain at least 4\,000 bound baryonic (star\,+\,gas) particles so that its internal kinematics are well resolved, including systems with $\mathrm{SFR}\simeq0$ that correspond to fully quenched galaxies. We also exclude any subhalo flagged with problems in its merger tree.
The complementary selection criteria outlined above yield a final sample of 7\,238 unique galaxies, covering:

\noindent $\cdot~10^{8.5} $$\,\leq\,$$ M_\star/\rm M_\odot $$\,\leq\,$$ 10^{11.6}$ (\small{Total mass of all bound stellar particles)}\\
$\cdot~0.0 \leq \rm SFR/(M_\odot\,yr^{-1}) \leq 54.1$ \small{(SFR averaged over the past 100\,Myr)}\\
$\cdot~10^{9.8} $$\,\leq\,$$ M_{200}/\rm M_\odot $$\,\leq\,$$ 10^{14.3}$ \small{(Total mass of all bound particles within $R_{200}$)} \normalsize

\noindent We show the distribution of our sample as a function of stellar mass, host halo mass, and SFR in Appendix\,\ref{AppendixA}.

\subsection{Environment and infall time algorithm}
\label{sec:environment}
To classify galaxies as centrals or satellites and to determine their accretion histories, we developed a merger-tree-based algorithm using the \textsc{Sublink} trees of the IllustrisTNG simulation. This approach follows the evolutionary track of each galaxy (hereafter subhalo) along its main progenitor branch and identifies the moment at which it first becomes a satellite within a larger system.

\begin{figure*}[htbp!]
  \centering
\includegraphics[width=1\textwidth]
{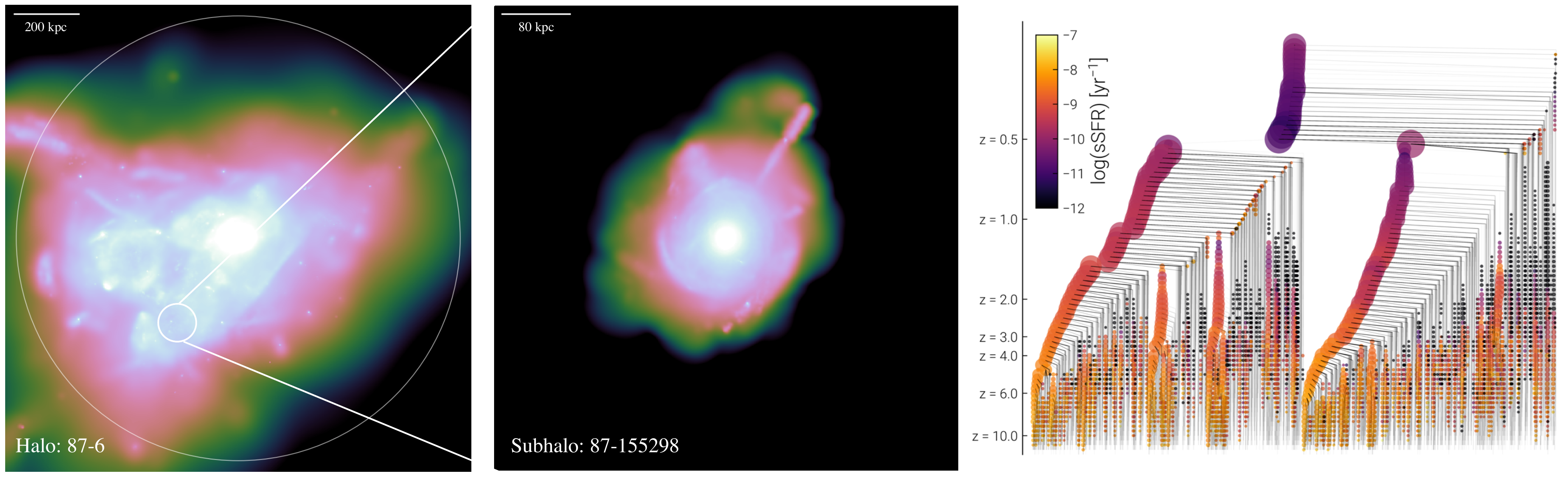}
    \caption{\textit{Left~panel:} Stellar density map of the TNG-50 mock halo 87-6 at $z$\,=\,0.15, where the subhalo 87-155298 (\textit{middle~panel}) is located in.
    \textit{Right~panel:} Corresponding merger tree of 87-155298 color-coded by the specific SFR. Each circle in the figure represents a past subhalo progenitor in the galaxy's history, with sizes proportional to their stellar mass at the corresponding snapshots.
    As shown in the \textit{right~panel}, at $z$$\,\sim\,$$0.5$ this subhalo underwent its last major merger, a nearly equal-mass event (stellar mass ratio ${\simeq}1{:}1.2$).
    Redshift increases from top (local Universe) to bottom (earlier epochs).}
\label{fig:merger_tree}
\end{figure*}
For each subhalo at $z$$\,\sim\,$$0$, we first identify its main progenitor branch by following the \texttt{MainLeafProgenitorID} through the merger tree. The index at which this identifier no longer changes corresponds to the main leaf progenitor, i.e. the earliest progenitor along the main evolutionary track. Starting from this object, we traverse the tree forward in time (towards lower redshift), tracking all its descendants.
Figure~\ref{fig:merger_tree} shows the example of halo 87-6 (\textit{left panel}) at $z$$\,\sim\,$$0.15$, which hosts the subhalo 87-155298 present in our sample (\textit{middle panel}). The \textit{right panel} of Fig.\,\ref{fig:merger_tree} displays the merger tree of subhalo 87-155298. Each circle represents a progenitor subhalo in the galaxy's evolutionary history, with sizes proportional to the stellar mass at the corresponding snapshot and colors indicating the specific SFR (sSFR).

At each snapshot, we compare the \texttt{SubhaloID} of the descendant to the \texttt{FirstSubhaloInFOFGroupID}, which identifies the central (most massive) subhalo of the corresponding FoF halo. When these two quantities are equal, the object is the central galaxy of its host halo. The transition from central to satellite is therefore defined as the first snapshot at which these identifiers differ, indicating that the subhalo has been accreted into a more massive system.

We define the \textit{first infall time} as the snapshot corresponding to this transition, i.e. the first time the galaxy becomes a satellite. Similarly, the \textit{last infall time} is defined as the most recent snapshot at which this transition occurs, allowing for the possibility of multiple infall events. In practice, we adopt a definition in which the infall snapshot corresponds to the first snapshot where the subhalo is already part of the host FoF group, ensuring consistency between first and last infall times in the case of a single accretion event.

This procedure provides a robust classification of galaxies into centrals (objects that have never undergone infall) and satellites (objects with at least one infall event), as well as a consistent estimate of their accretion histories. Similar merger-tree-based approaches have been used in the literature to study subhalo evolution and environmental effects in IllustrisTNG \cite[e.g.,][]{Donnari+21,Donnari+21b,Heinze+24}.

\subsection{Gas temperature and masses}\label{hotgasmass}
The total gas mass of each subhalo was computed by summing the masses of all gas cells gravitationally bound to the subhalo, as identified by the \textsc{Subfind} algorithm in TNG50. 
The gas temperature was estimated from the internal energy of each gas cell assuming an ideal monoatomic gas. Following the IllustrisTNG formalism, the temperature was computed as

\begin{equation}
T = (\gamma - 1)\,\frac{u}{k_{\rm B}}\,\mu,
\end{equation}
where $\gamma = 5/3$ is the adiabatic index, $u$ is the internal energy per unit mass, $k_{\rm B}$ is the Boltzmann constant, and $\mu$ is the mean molecular weight. The latter was calculated using the hydrogen mass fraction $X_{\rm H}=0.76$ and the electron abundance, $x_{\rm e}$, provided by the simulation:
\begin{equation}
\mu = \frac{4\,m_{\rm p}}{1 + 3X_{\rm H} + 4X_{\rm H}x_{\rm e}},
\end{equation}
where $m_{\rm p}$ is the proton mass.
In this work, we define hot gas as any gas cells above $10^5$\,K, and cool gas as all gas cells below $3\times10^4$\,K\footnote{Although ``cool'' and ``cold'' gas do not refer to the same phase, cosmological hydrodynamical simulations do not resolve the cold molecular gas ($T \lesssim 50$\,K) or the cold neutral atomic (HI) medium ($T \lesssim 100$\,K).}.

\section{Kinematic analysis}\label{Sec3}
\subsection{The \textsc{Kinemetry} package}\label{Kinemetry}
To identify kinematically asymmetric galaxies, we use the \textsc{Kinemetry} package,\footnote{The package is available at: \url{https://www.aip.de/en/members/davor-krajnovic/kinemetry/}} developed by \cite{Kinemetry}.
This code analyzes 2D maps of the moments of the line-of-sight velocity distribution (intensity, velocity, velocity dispersion) by performing harmonic expansions along best-fitting ellipses, providing a robust quantification of kinematic structures and subcomponents \citep{Kinemetry}. It models velocity profiles as a Fourier series,
\begin{equation}\label{eq:kinemetry}
    K(a,\psi)=A_0(a)+\sum\limits_{n=1}^{N} A_n(a)\sin(n\psi)+B_n(a)\cos(n\psi),
\end{equation}
where $\psi$ is the eccentric anomaly, $a$ the semi-major axis length, and $A_0$ the systemic velocity. Outputs include the position angle, ellipticity, and asymmetry coefficients. 
The final ellipse parameters obtained by minimization are then used to describe an elliptical ring from which a kinematic profile is extracted and expanded on to the harmonic series of Eq. (\ref{eq:kinemetry}), where the coefficients ($A_n$, $B_n$) are determined by a least-squares fit with a basis $\{1, \cos(\psi), \sin(\psi), ... , \cos(N\psi), \sin(N\psi)\}$ \cite[][]{Kinemetry}.
\textsc{Kinemetry} has been widely validated as a tracer of kinematic asymmetries, successfully identifying features linked to mergers, gas stripping, bars, and feedback-driven turbulence \citep[e.g.,][]{Shapiro+08, Liu+13, Holmes+15, Bloom+18}. We quantify asymmetries using the parameter
\begin{equation}\label{eq:asym}
    \overline{I_{\rm asym}}=\overline{\left(\frac{k_2+k_3+k_4+k_5}{4\,k_1}\right)},
\end{equation}
where $k_1$ is the amplitude of the rotating component and higher-order terms ($k_2$–$k_5$) represent nonrotating contributions \citep[][]{Kinemetry, Shapiro+08}. 
While some studies consider only odd modes \cite[e.g., $k_3$ and $k_5$;][]{Bloom+17, Bloom+18}, we include all terms to also capture signatures of major mergers. Equal contributions from rotating and nonrotating components yield $I_{\rm asym} $$\,\geq\,$$ 0.25$ ($\log_{10} I_{\rm asym} $$\,\geq\,$$ -0.60$). \citet{Bloom+17} adopted a threshold of $I_{\rm asym} $$\,\geq\,$$ 0.065$ ($-1.19$ in $\log_{10}$), while \citet{Feng+22} used $I_{\rm asym} $$\,\geq\,$$ 0.039$ ($-1.41$ in $\log_{10}$). Here, following \cite{deIsidio+26}, we adopt an intermediate threshold of $I_{\rm asym} $$\,\geq\,$$ 0.05$ ($-1.30$ in $\log_{10}$), corresponding to a $\,\geq\,$18\% nonrotating contribution.

\subsection{Gas kinematics in TNG50}
\begin{figure*}[htbp!]
  \centering
\includegraphics[width=0.9\textwidth]
{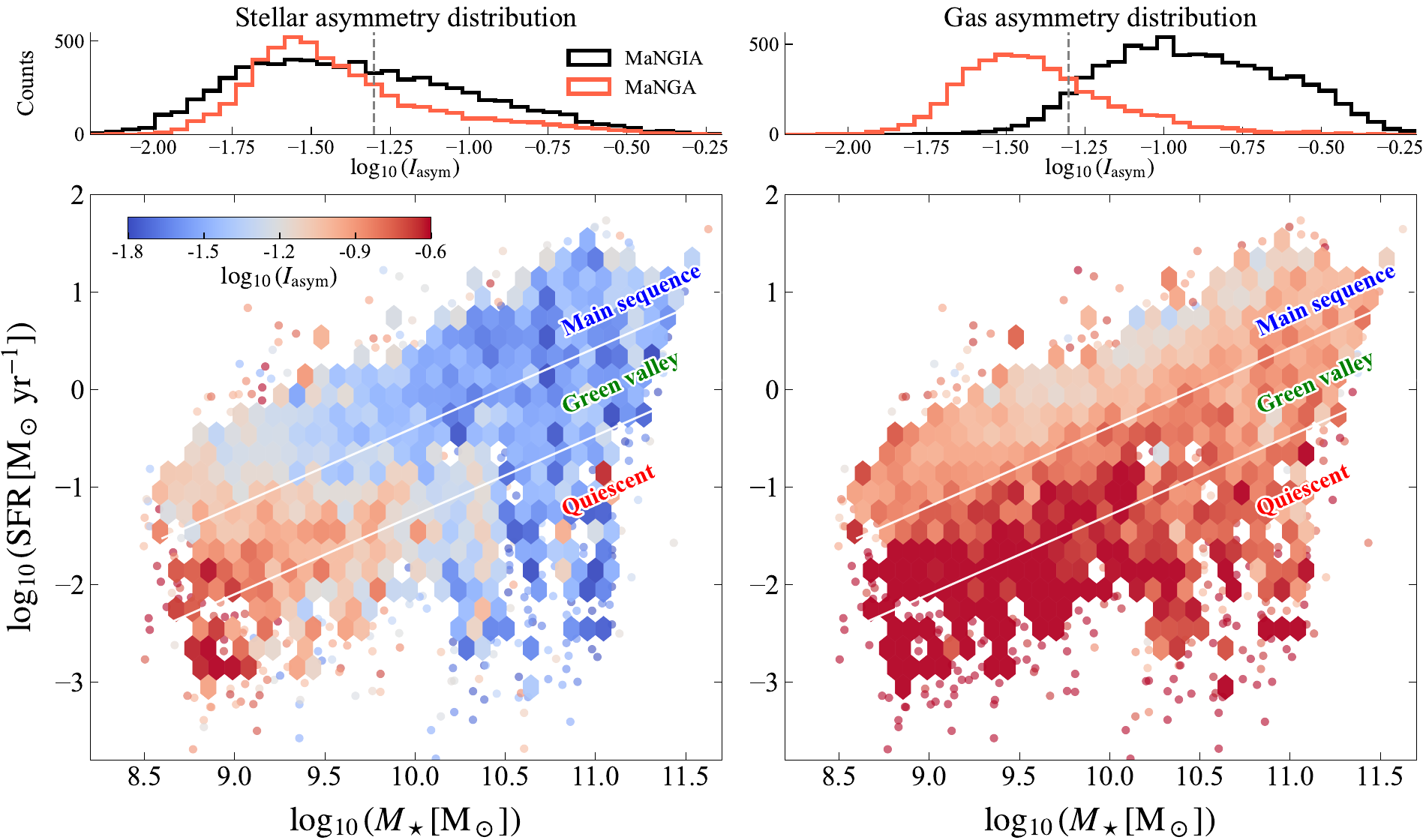}

    \caption{SFR--$M_\star$ plane of our mock MaNGA sample, color-coded by the degree of kinematic asymmetry, $I_{\rm asym}$ (Eq.~\ref{eq:asym}), of the stellar (\textit{left panel}) and gas (\textit{right panel}) velocity fields.
    Following our criterion, kinematically asymmetric galaxies are those with $\log_{10} I_{\rm asym}$$\,\geq\,$$-1.30$ (i.e. $I_{\rm asym}$$\,\geq\,$$0.05$).
    The upper panels show the distribution of galaxies as a function of $\log_{10} I_{\rm asym}$ for both the MaNGA and mock MaNGA samples.
    The vertical dashed line separates galaxies into kinematically symmetric and asymmetric systems.
    Although the stellar component of the mock sample is broadly consistent with the distribution of asymmetries measured in the observational MaNGA sample presented by \cite{deIsidio+26}, the gas distribution is significantly shifted toward higher degrees of asymmetry. We therefore do not consider gas velocity maps in our analysis, as direct comparisons with the observed MaNGA gas kinematics would likely lead to misleading conclusions.
    The lines indicate the boundaries separating star-forming, green valley, and quiescent galaxies, as defined by \cite{Behroozi+19}.
    The main panels use hexagonal binning, with 35 hexagons spanning the horizontal axis; each hexagon is color-coded by the mean of $\log_{10} I_{\rm asym}$ over the galaxies it contains, and is drawn only where $\geq$3 objects occupy the same region of the plane. Galaxies falling in less populated hexagons are shown individually as points.
}
\label{fig:Asymmetries}
\end{figure*}

We apply the \textsc{Kinemetry} package to both stellar and gas velocity maps of all mock galaxies. 
However, when compared to the observed MaNGA sample of \citet{deIsidio+26} (see Fig.\,\ref{fig:Asymmetries}), a clear discrepancy emerges: while the stellar asymmetry distributions are broadly consistent (59.0\% of MaNGIA and 72.1\% of MaNGA galaxies are symmetric, with an overlap coefficient of 84\%), the gas kinematics are not. In fact, only 6.3\% of MaNGIA galaxies have symmetric gas velocity fields, compared with 71.2\% in MaNGA, an order-of-magnitude deficit.
The upper panels of Fig.\,\ref{fig:Asymmetries} compares the asymmetry distributions derived from the stellar and gas velocity maps for our mock MaNGA sample and for the observed MaNGA sample analyzed in \cite{deIsidio+26}.

This mismatch likely reflects limitations in both the modeling of the gas component and the forward-modeling procedure. Feedback is implemented through subgrid prescriptions that regulate energy and momentum injection below the resolution scale (e.g., \citealt{Weinberger+17a,Weinberger+17,Hayward+17}), and may contribute to enhanced turbulence or irregular gas motions.
In addition, the mapping between simulated gas cells and H$\alpha$ emission is approximate, as the multiphase ISM is not fully resolved in cosmological simulations, also shown by the differences in gas properties from smaller-scale higher resolution simulations \cite[e.g.,][]{Tacchella+22c,McClymont+24,Hidalgo-Pineda+26}. 
The higher intrinsic resolution of TNG50 also preserves small-scale fluctuations that are partially smoothed out in real IFU data but not entirely suppressed in mock observations \cite[e.g.,][]{Barrera-Ballesteros+15}.

Further uncertainties arise from the forward-modeling, which cannot fully reproduce observational systematics such as beam smearing, line-fitting uncertainties, and correlated noise \citep{Westfall+19}. Moreover, environmental processes affecting the gas (e.g., ram pressure stripping) depend sensitively on the structure of the CGM and IGrM/ICM, whose cold and dense phases are not well converged in cosmological simulations and can vary significantly with resolution \cite[e.g.,][]{Tonnesen+09,Bahe+13,Peeples+19}. This may affect the predicted level of gas disturbances in simulated galaxies.

Nevertheless, we retain information on the gaseous component by considering global gas properties of the subhalos, including the total gas mass in different temperature phases (hot, $T $$\,\geq\,$$ 10^{5}\,$K, and cool, $T $$\,\leq\,$$ 3 \times 10^{4}\,$K), as well as the gas temperature, internal energy, electron abundance, and density. These quantities offer a more reliable characterization of the gas content, being less sensitive to the modeling and observational uncertainties that affect spatially resolved gas kinematics.

\section{Results}
\label{results}
We present our results in this section. For clarity, we organize our main findings in the subsections below.

\subsection{Kinematic transformation during quenching phase}\label{symmetric}
Figure~\ref{fig:Asymmetries} shows the SFR--$M_{\star}$ plane color-coded by the stellar kinematic asymmetry parameter measured as described in $\S$\ref{Kinemetry} for all galaxies in our sample. 
A clear trend emerges in which the vast majority of galaxies with disturbed stellar kinematics are preferentially located at low stellar masses, particularly below $M_{\star}$$\,\sim\,$$10^{10}\,{\rm M_{\odot}}$.
At higher stellar masses ($M_{\star}$$\,\gtrsim\,$$10^{10.5}\,{\rm M_{\odot}}$), the stellar kinematics are predominantly regular and rotation-supported, even among galaxies transitioning through the green valley or already located in the quenched region.

This behavior is qualitatively consistent with the observational results presented by \citet{deIsidio+26}, who found that most galaxies in the local Universe exhibit symmetric stellar kinematics independently of their star formation activity. In particular, the observational analysis showed that the satellite region of the SFR--$M_{\star}$ plane is strongly dominated by galaxies without significant stellar asymmetries, suggesting that the dominant environmental quenching pathways do not strongly perturb the stellar kinematic structure of galaxies. Our TNG50 sample reproduce this overall trend reasonably well, especially at the low-mass end where asymmetric galaxies are more frequently found.

However, an important discrepancy emerges at high stellar masses. While \citet{deIsidio+26} identified a non-negligible population of massive quenched galaxies with disturbed stellar kinematics---interpreted as signatures of dry mergers, tidal interactions, or past dynamical perturbations---TNG50 appears to significantly underproduce these systems \cite[see also][]{Sarmiento+23}. In the simulation, massive galaxies above $M_{\star}$$\,\gtrsim\,$$10^{10.5}\,{\rm M_{\odot}}$ are overwhelmingly dominated by regular stellar velocity fields, suggesting that the mechanisms producing strong stellar disturbances in massive quenched galaxies (sSFR\,$\leq10^{-11}\,$yr$^{-1}$) may be less efficient, less frequent, or more rapidly relaxed in TNG50 than in the observed Universe.

The prevalence of regular stellar kinematics among star-forming galaxies is also particularly noteworthy. Along the star-forming main sequence, most galaxies exhibit ordered rotation with little evidence for strong stellar disturbances. This suggests that a significant fraction of galaxies may evolve toward quenching without undergoing major stellar kinematic transformation \cite[e.g.,][]{Cortese+19,deIsidio+26}. In such a scenario, quenching would primarily proceed through processes that gradually suppress star formation while largely preserving the stellar dynamical structure, such as halo gas depletion, or long-timescale feedback maintenance. This interpretation is broadly consistent with the observational picture proposed by \citet{deIsidio+26}, in which the dominant quenching pathways leave minimal stellar kinematic imprints by the time galaxies become fully quenched.

\subsection{Gas evolution and timescales for CGM depletion}
\label{hot_gas}
One of the most effective ways to quench a galaxy is to suppress the continuous replenishment of its cold gas reservoir. In this scenario, star formation does not cease immediately through the direct removal of the star-forming gas in the disk, but instead declines over longer timescales as the galaxy exhausts the remaining fuel available for star formation. 

\begin{figure*}[htbp!]
  \centering
\includegraphics[width=0.99\textwidth]
{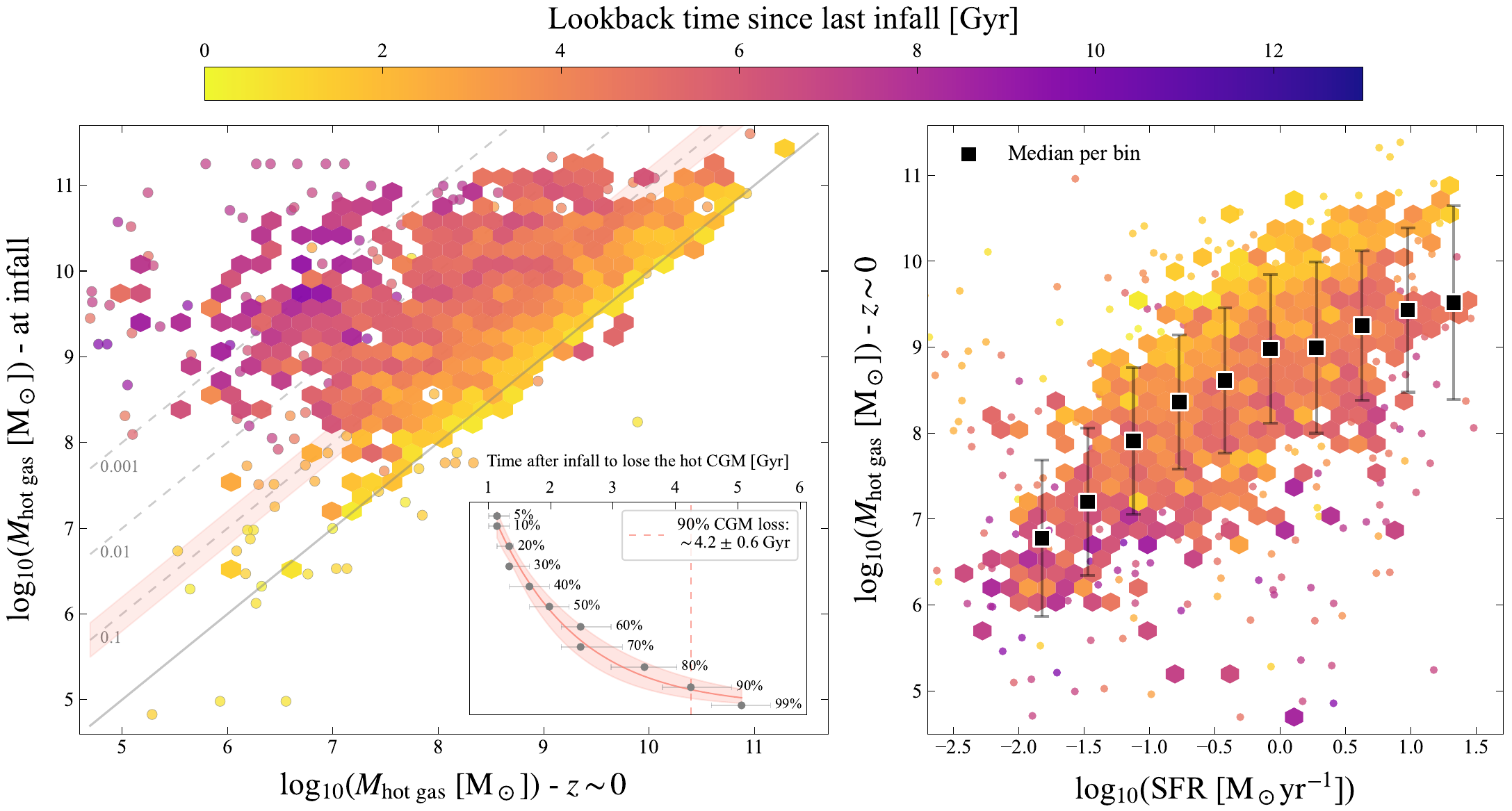}
    \caption{\textit{Left panel}: Hot gas mass of satellites at the time of infall into their host halo versus their current hot gas mass, color-coded by the lookback time since their last infall. The gray solid line represents the 1:1 relation, while the dashed lines mark present-day masses equal to 10\%, 1\%, and 0.1\% of the infall mass.
    The red shaded area marks the region used to calculate the uncertainty on the 90\% decrease shown in the inset panel. 
    \textit{Right panel}: Hot gas mass as a function of SFR, also color-coded by the lookback time since last infall. Both panels reveal a clear trend between hot gas depletion and infall time, with galaxies that fell into their host halos earlier exhibiting the most depleted hot gas reservoirs.
    In the inset plot in the lower-right corner of the \textit{left panel}, we group satellites by the fraction of their infall hot gas that they currently retain: for each target fraction we select all satellites whose present-day $M_{\rm hot\,gas}/M_{\rm hot\,gas,\,infall}$ lies within $0.1$\,dex of that value, and compute the median lookback time since their last infall. Each point therefore reports how long ago satellites at a given depletion level were accreted.
    Satellites currently retaining ${\sim}50\%$ of their infall hot CGM were accreted a median of ${\sim}2$\,Gyr ago, while those retaining only ${\sim}10\%$ were accreted $4.2^{+0.6}_{-0.6}$\,Gyr ago, indicating a strong link between hot CGM depletion and time since infall. Error bars represent $3\sigma$ bootstrap uncertainties on
    the median lookback times.
}
    \label{fig:MainPlot}
\end{figure*}

To further investigate the role of the CGM in shaping the quenching pathways of galaxies, we perform a detailed analysis of the hot gas content ($T $$\,\geq\,$$ 10^{5}$\,K) of our sample. For all galaxies, we compute the present-day hot and cool gas masses as described in $\S$\ref{hotgasmass}, while for satellites we additionally reconstruct the evolution of the gas reservoir along the merger tree. 
Using an interpolation algorithm applied to the subhalo histories, we compute both hot and cool gas masses at key evolutionary stages, including the epoch of the infall into the current host halo and the time when the galaxy first became a satellite of a larger halo. This approach allows us to directly connect the depletion of the CGM with orbital history and environmental processing.

The left panel of Fig.\,\ref{fig:MainPlot} compares the current hot gas mass of our satellite population ($\sim$2\,800 galaxies) with the hot gas mass these systems had at the time of infall into their present-day host halo. A clear trend is observed: nearly all satellites exhibit substantial depletion of their hot gas reservoirs after infall. 
More importantly, the majority of satellites have already lost more than $\sim$60\% of their original hot CGM content by the present epoch. This result provides direct evidence that the environmental impact on satellite galaxies is not restricted to the cold gas within the disk, but instead acts efficiently on the surrounding gas reservoir that sustains long-term star formation.

When comparing the infall times of currently quenched satellites to those of satellites with ongoing star formation, we find that the mean infall time of non-quenched satellites is significantly shorter than that of quenched ones. 
Specifically, we find that the average infall time for satellites with ongoing star formation is $4.3^{+0.3}_{-0.3}$\,Gyr, whereas for quenched satellites, this value is significantly higher at $6.5^{+0.3}_{-0.3}$\,Gyr.
This result suggests that satellites tend to continue forming stars for about at least 3\,Gyr after infall before environmental effects completely shut down their star formation.

When combining the hot gas measurements with infall times, an even more striking picture emerges: for the first time, we present a clear time evolution of hot CGM depletion after infall, with the depletion rate strongly correlated with the time elapsed since the satellite entered its current host halo.
The color gradient shown in Fig.\,\ref{fig:MainPlot} demonstrates that satellites residing for longer periods within their host environments systematically exhibit larger hot gas losses.
Importantly, this time dependence is not driven by galaxy stellar mass: measuring the CGM-depletion timescale in separate bins of $z$$\,\sim\,$$0$ stellar mass yields a time to lose $90\%$ of the hot gas that is consistent across the full mass range (see Appendix\,\ref{AppendixB}; Fig.\,\ref{fig:Mstar_dependence}), with stellar mass setting only the overall normalization of the hot-gas reservoir rather than the rate at which it is stripped.
This behavior strongly supports a progressive environmental depletion scenario rather than a purely impulsive, short quenching phase, and confirms that the observed trends are governed by environment rather than by galaxy mass.

The inset panel in Fig.\,\ref{fig:MainPlot} quantifies this evolution by showing the median timescales required for satellites to lose different amounts of their original hot gas reservoirs. We find that satellites lose on average $\sim$50\% of their initial hot CGM within $\sim$2\,Gyr after infall. 
Interestingly, this timescale coincides with the expected interval between infall and first pericentric passage in group and cluster environments, typically of order $\sim$1--4\,Gyr depending on host mass and orbital properties.
This connection strongly suggests that environmental processes become particularly effective near first pericentric passage, where ram pressure, tidal forces, and interactions with the dense IGrM/ICM reach maximum efficiency \cite[e.g.,][]{Lotz+19}.

To establish whether the long hot-gas depletion timescale reflects genuine starvation rather than the direct removal of the star-forming fuel, we repeat the analysis of Fig.\,\ref{fig:MainPlot} for the cool gas component ($T $$\,\leq\,$$ 3\times10^{4}$\,K). The result is shown in Appendix\,\ref{AppendixC} (see Fig.\,\ref{fig:coolgas}).
While many satellites lose cool gas between infall and the present day, a substantial fraction instead increase their cool-gas mass over the same interval. The decline of the hot CGM seen in Fig.\,\ref{fig:MainPlot} can therefore not be attributed solely to hot gas removal: part of it corresponds to hot gas cooling onto the disk, where it replenishes the cool reservoir.

The cool gas is itself depleted only on long timescales. We measure a median time of $\sim$$4.9^{+0.9}_{-0.9}$\,Gyr to remove 90\% of the cool gas, comparable within 1$\sigma$ to the value found for the hot CGM.
The depletion is slow at early times---less than 10\% of the cool gas is lost within the first $\sim$2\,Gyr after infall---which is inconsistent with the rapid ($\lesssim$1\,Gyr) removal expected from cool gas stripping processes (e.g., ram-pressure stripping). 
Most of the decline (from 40\% to 90\%) indeed occurs abruptly, within $\sim$1\,Gyr, but only after $\sim$4\,Gyr have elapsed since infall. Furthermore, for each satellite we estimate the gas-depletion timescale expected from star formation alone, $\tau_{\rm dep} = M_{\rm gas}/{\rm SFR}$, evaluated at infall from its cool-gas content and SFR at that epoch. We find a median $\tau_{\rm dep} $$\,\approx\,$$ 5.5^{+0.2}_{-0.1}$\,Gyr, comparable to the timescales over which both the hot and cool gas reservoirs are observed to decline.

The right panel of Fig.\,\ref{fig:MainPlot} shows a positive correlation between the present-day hot gas content of satellites and their current SFR, steeper at low $M_{\rm hot~gas}$ and flattening toward higher hot gas masses.
Satellites with earlier infall times preferentially populate the low-$M_{\rm hot~gas}$, low-SFR region, while recently accreted systems retain both larger hot reservoirs and higher SFRs. At high SFR the relation shows a large scatter, spanning nearly two orders of magnitude in hot gas mass, part of which correlates with infall time.

For comparison, we show in Fig.\,\ref{fig:centrals} the same plane for central galaxies, color-coded by the specific SFR. Centrals separate into two regimes: star-forming systems (sSFR>$10^{-11}\,$yr$^{-1}$) define a positive $M_{\rm hot~gas}$--SFR sequence analogous to that of satellites, while quenched centrals occupy the low-SFR region while retaining large hot reservoirs ($M_{\rm hot~gas}$$\,\sim\,$$ 10^{10-11}\,\rm M_\odot$), comparable to or larger than those of actively star-forming systems. In contrast to satellites, low star formation in centrals is therefore not accompanied by a depletion of the hot CGM.

\subsection{Mass accretion/loss history during quenching phase}\label{MassEvolution}

\begin{figure*}[htbp!]
  \centering
\includegraphics[width=0.98\textwidth]
{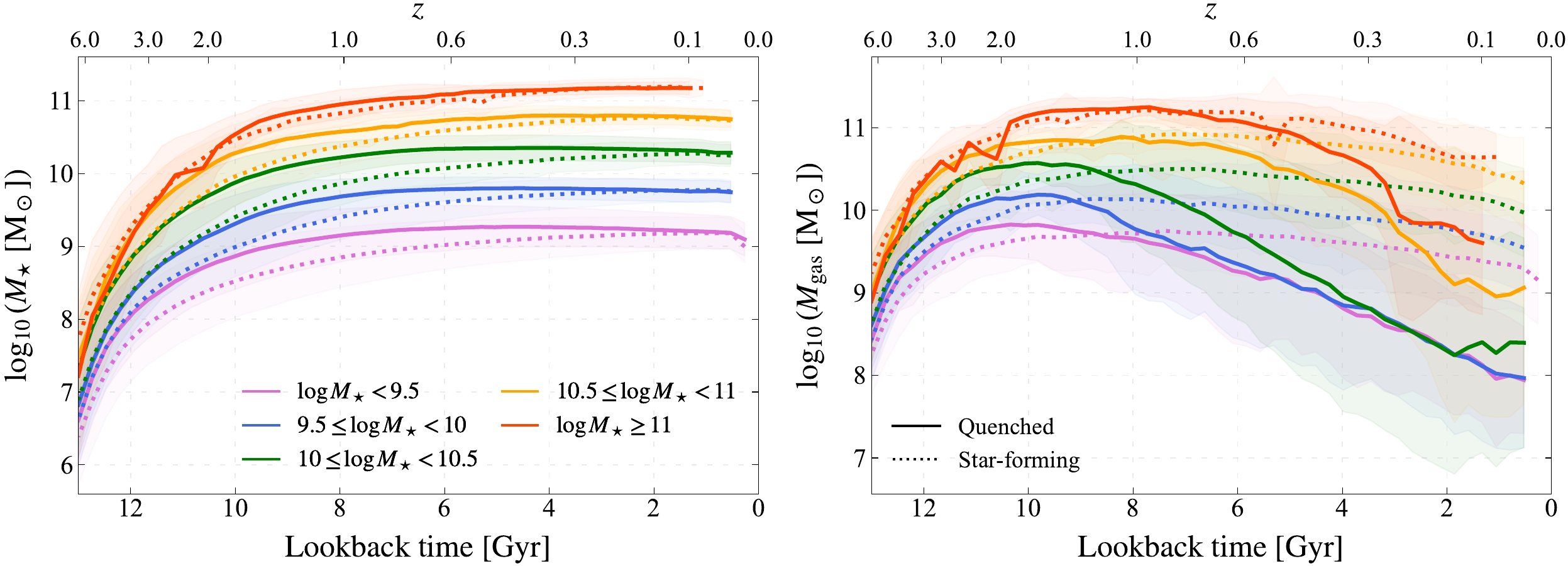}
    \caption{\textit{Left~panel:} Stellar mass histories of our satellite subsample. Curves show the mean stellar mass of the galaxies contributing to each bin, while the shaded areas indicate the $1\sigma$ dispersion.
    Solid curves represent present-day quenched (${\rm sSFR}$$\,\leq\,$$10^{-11}\,\rm yr^{-1}$) satellites, whereas dotted curves show galaxies that are currently star-forming.
    To investigate trends with stellar mass, we split the sample into five stellar mass bins based on their present-day stellar masses ($z$$\,\sim\,$$0$).
    \textit{Right panel}: Same as the \textit{left panel}, but showing the evolution of the total gas mass.
    Although the stellar mass accretion histories evolve relatively smoothly over time regardless of whether satellites are currently quenched or star-forming, the evolution of the gas content depends strongly on the present-day star formation state, particularly in the lower-mass bins ($M_\star$$\,\leq\,$$10^{10.5}\,\rm M_\odot$).
    The median total gas mass of all satellites declines after infall into a more massive halo, although the fraction of gas lost depends strongly on whether the galaxy is currently quenched.
    While the stellar component appears to be only weakly affected by environmental processing after infall, the gas component is efficiently removed, consistent with the results presented in Fig.\,\ref{fig:MainPlot}.
    Curves are shown only where at least 50 subhalos contribute at the corresponding
    snapshot.
}
    \label{fig:MassEvolution}
\end{figure*}

In Figure~\ref{fig:MassEvolution}, we investigate the stellar- and gas-mass evolution of satellite galaxies as a function of lookback time, separating the population into present-day stellar mass bins and current star-forming versus quenched systems.

The stellar mass evolution shown in the left panel shows that satellite galaxies, on average, experience little to no significant stellar mass loss throughout their evolution. Instead, the main difference between present-day quenched and star-forming satellites lies in the epoch at which stellar mass growth slows down or effectively ceases. 
Present-day quenched satellites systematically reach their maximum stellar masses substantially earlier than star-forming satellites of the same present-day stellar mass bin. After this early growth phase, their stellar mass evolution becomes considerably flatter, indicating that these galaxies assembled most of their stellar content early and subsequently experienced only limited additional stellar mass growth. In contrast, present-day star-forming satellites continue increasing their stellar masses over much longer timescales, maintaining sustained star formation down to recent epochs.

We also note that the magnitude of this offset depends on stellar mass. Lower-mass satellites ($M_\star $$\,\lesssim\,$$ 10^{10.5}\,\rm M_\odot$) exhibit a larger offset between the stellar mass histories of quenched and star-forming galaxies than their high-mass counterparts, while the difference progressively decreases toward higher stellar masses.
For satellites with present-day stellar masses above $\sim$$10^{10.5}\,\mathrm{M_\odot}$, the stellar mass assembly histories of quenched and star-forming systems become significantly more similar. 
This behavior indicates that environmental quenching leaves a stronger imprint on the stellar mass assembly histories of low-mass satellites rather than on massive systems.

We also analyze the gas mass evolution of satellites, as shown in the right panel of Fig.\,\ref{fig:MassEvolution}. Independent of present-day stellar mass or star formation state, essentially all satellites lose gas over cosmic time. 
However, the magnitude and time evolution of this depletion differ dramatically between quenched and star-forming systems. Present-day star-forming satellites exhibit comparatively flat gas loss histories after reaching their peak gas content, indicating that although some gas depletion occurs, these galaxies retain sufficient gas reservoirs to sustain star formation over extended periods. 
In contrast, present-day quenched satellites show significantly steeper declines in gas mass following their peak gas content, suggesting that quenching is closely associated with accelerated gas depletion.

An interesting result is observed when comparing the gas mass evolution with the average satellites' first infall times.
Across all stellar mass bins, the average epoch at which galaxies first become satellites closely coincides with the moment when their gas content begins to decrease significantly. 
The difference in the gas loss history between quenched and star-forming satellites is particularly pronounced in the lowest stellar mass bins. Satellites with present-day stellar masses below $\sim$$10^{10.5}\,\mathrm{M_\odot}$ exhibit the most dramatic gas losses, with their gas reservoirs declining rapidly after reaching peak gas mass. On average, these galaxies lose $47^{+10}_{-6}\%$ of their gas mass within $\sim$2\,Gyr after infall, strongly suggesting that they experience one or multiple episodes of environmental stripping (either via hydrodynamical or gravitational processes) throughout their evolution. 
In contrast, satellites in the higher stellar mass bins ($M_\star $$\,\gtrsim\,$$ 10^{10.5}\,\mathrm{M_\odot}$) lose $\sim$30\% of their gas reservoirs over the same period, indicating that environmental gas removal is substantially less efficient in the most massive systems, with the stripping efficiency being nearly a factor of two lower than in their low-mass counterparts.
This behavior is fully consistent with the expectation that low-mass systems are more susceptible to efficient environmental stripping due to their shallow potential wells and weaker gravitational restoring forces. 

\subsection{Metallicity evolution during quenching phase}\label{subsec:metallicity}
It has been shown that satellite galaxies exhibit enhanced stellar metallicities when compared to centrals at fixed stellar mass, particularly at low masses \cite[e.g.,][]{Pasquali+10,Wetzel+13,Peng+15,Gallazzi+21}. 
In a quenching phase driven by starvation\footnote{\cite{Peng+15} opted to use the term "strangulation", although both terms (\textit{starvation}, \textit{strangulation}) refer to the same physical process.} as described in \citet{Peng+15}, the removal of cosmological gas accretion prevents the replenishment of the interstellar medium with low-metallicity gas, while star formation continues consuming the remaining gas reservoir over several Gyr. 
As a consequence, the stellar metallicity of satellites progressively increases. 
In this framework, \cite{Peng+15} claimed that the metallicity offset between centrals and satellites is expected to be more pronounced at low stellar masses because satellites are expected to suffer more from starvation than centrals.

\begin{figure*}[htbp!]
  \centering
\includegraphics[width=1\textwidth]
{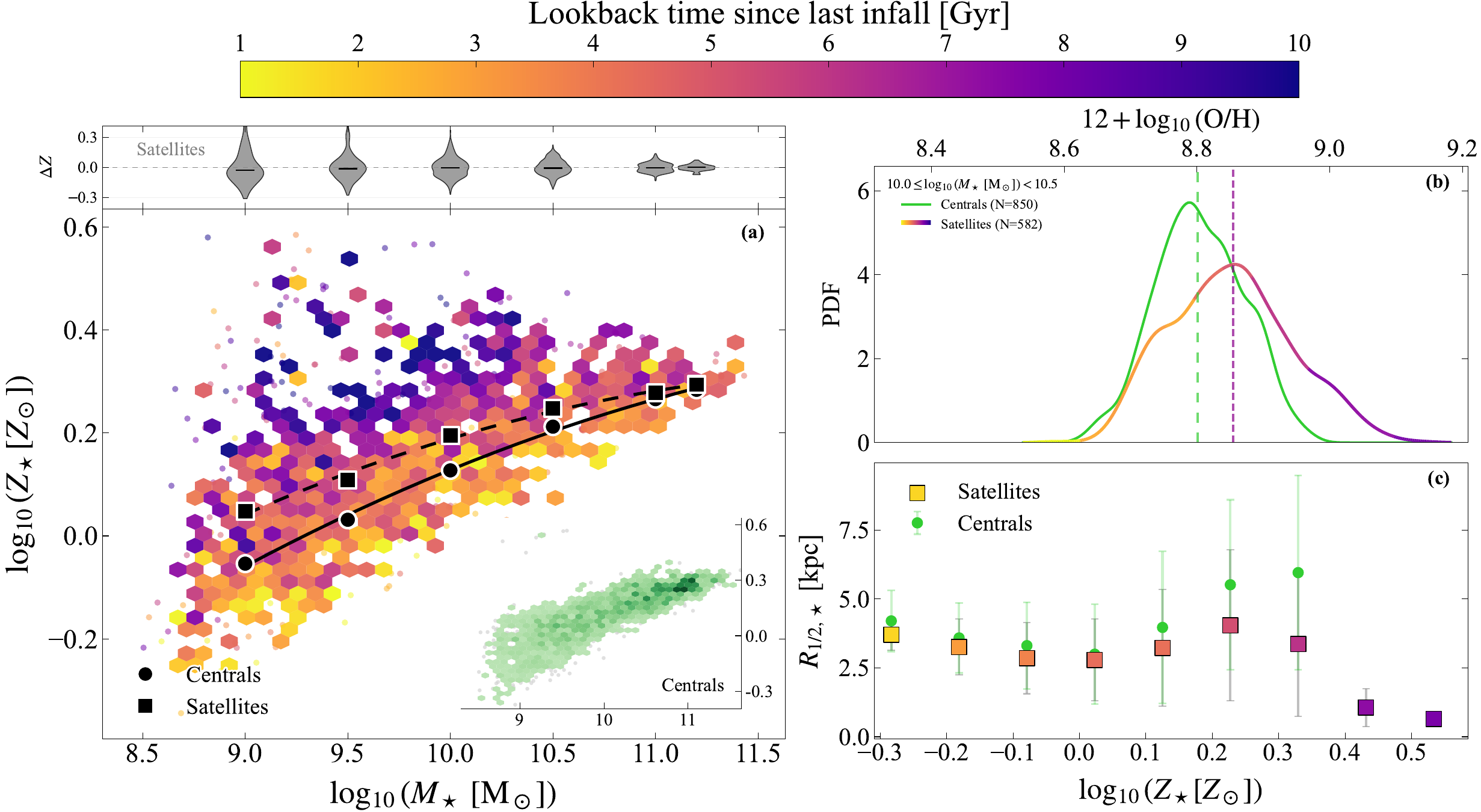}
    \caption{\textit{Panel~a:} Mass--metallicity (stellar) relation of our satellite subsample, color-coded by the lookback time since last infall.
    For comparison, we display the mean relations for both centrals (circles) and satellites (squares).
    The inset panel shows the same relation for the central subsample, with colors representing the density.
    \textit{Panel~b:} Probability density function of the stellar metallicity for a representative stellar mass bin ($10^{10.0} $$\,\leq\,$$ M_\star/\rm M_\odot $$\,<\,$$ 10^{10.5}$), comparing satellites (colored curve) and centrals (green curve).
    \textit{Panel~c:} Size--metallicity relation of our sample. As in the previous panels, satellites are color-coded by the lookback time since last infall.
    $R_{1/2,\,\star}$ corresponds to the half stellar mass radius of the galaxy.
    The systematically higher metallicities and larger scatter of satellites (shown in the violin plot above \textit{panel~a}) relative to centrals, together with the structural compaction in the satellite population, are consistent with preferential loss of extended, relatively metal-poor gas, although starvation and differences in assembly history may also contribute. 
    }
    \label{fig:metallicity}
\end{figure*}

IllustrisTNG reproduces this general trend. Satellites display systematically higher metallicities than centrals at fixed stellar mass, with the metallicity offset becoming progressively smaller toward higher masses and disappearing at $M_\star \gtrsim 10^{11}\,\rm M_\odot$ (see Fig.\,\ref{fig:metallicity}a).
Additionally, satellites exhibit a remarkably broad metallicity distribution at $M_\star $$\,\lesssim\,$$ 10^{10.5}\,\rm M_\odot$. In contrast, centrals display a much smoother and narrower metallicity sequence, with most systems remaining below $Z_\star/ Z_\odot$$\,\sim\,$$0.1$\footnote{$[12+\log_{10}({\rm O/H})]_\odot$$\,\approx\,$8.623, assuming $Z_\odot=0.0142$ \citep[][]{Asplund+09}.} at low stellar masses, mostly driven by the time from last infall. Indeed, the larger the spread in the time from the last infall, the larger the metallicity scatter at fixed stellar mass. As more massive galaxies exhibits a smaller spread in time from last infall, they consistently show also a smaller scatter in metallicity. 

To further investigate this effect, we compare representative samples of centrals and satellites in the stellar mass range $10^{10.0}$$\,\leq$$\,M_\star/\rm M_\odot < 10^{10.5}$. The corresponding probability density functions are shown in Fig.\,\ref{fig:metallicity}b. As expected, satellites display systematically higher median metallicities ($12+\log_{10}({\rm O/H}) \approx 8.85$) than centrals ($\approx$8.80). 
However, the most important feature is the extended high-metallicity tail present exclusively in the satellite population. While galaxies with time from the last infall shorter than 3 Gyr behave just like centrals, satellites with a much larger time from last infall are significantly metal richer.

\section{Discussion}
\label{Sec5}
We present below a discussion of our main results, linking them to the quenching scenarios that most plausibly explain the observed trends.
For clarity, we separate and discuss each of the results presented in $\S$\ref{results} in the subsections below.

\subsection{The CGM as the primary regulator of satellite quenching}
\label{rapidthendelayed}
A central result of this work is the tight coupling between the hot CGM and the star formation activity of satellites. As we show in the previous section, according to IllustrisTNG the present-day hot gas mass correlates strongly with SFR: satellites with earlier infall times preferentially occupy the low-SFR, low-CGM-mass regime, and the degree of hot-halo depletion increases monotonically with environmental residence time (Fig.\,\ref{fig:MainPlot}).

This behavior is naturally understood if the CGM acts as the long-term fuel reservoir of galaxies \citep[e.g.,][]{Tumlinson+17,Davies+20,Peroux+20,Cortese+21}. 
In centrals, sustained cooling from the hot halo continually replenishes the cold, star-forming gas, allowing star formation to persist over cosmological timescales. For satellites, infall disrupts this cycle: once the circumgalactic reservoir is stripped or prevented from cooling, the disk can no longer be resupplied, and star formation declines as the residual cold gas is consumed. The steep low-$M_{\rm hot}$ end of the $M_{\rm hot}$--SFR relation (Fig.\,\ref{fig:MainPlot}) is the expected fingerprint of this regime, where the SFR becomes limited by the dwindling hot reservoir rather than by the
instantaneous cold-gas content.

At fixed SFR the relation shows substantial scatter, reaching nearly two orders of magnitude in hot gas mass at the high-SFR end. This scatter is well organized by infall time: at a given SFR, recently accreted satellites retain systematically larger hot reservoirs than those that fell in earlier (Fig.\,\ref{fig:MainPlot}). Stellar mass does not appear to set the rate of depletion. In fact, we show in Appendix\,\ref{AppendixB} that the time required to remove 90\% of the hot CGM is consistent across stellar-mass bins. Rather, stellar mass most plausibly enters through the distribution of infall times: more massive satellites tend to be accreted more recently, having spent much of their history as centrals of their own halos, and therefore preferentially populate the high-$M_{\rm hot}$, high-SFR end of the relation, whereas lower-mass satellites accreted at earlier epochs have had have had more time to deplete their gas reservoirs.

This temporal sequencing is consistent with a ``delayed-then-rapid'' quenching scenario \citep{Wetzel+13,Bahe+15}, in which satellites continue forming stars for several Gyr after infall before declining sharply. Our cool-gas analysis (Appendix\,\ref{AppendixC}) reproduces this two-phase behavior: less than 10\% of the cool gas is removed during the first $\sim$2\,Gyr after infall (the delayed phase)---and cooling from the still-massive hot halo can even grow the cool reservoir in a fraction of satellites---after which the reservoir drops abruptly, from 40\% to 90\% depletion within $\sim$1\,Gyr around $\sim$4\,Gyr post-infall (the rapid phase), with an e-folding time comparable to the $\lesssim$0.8\,Gyr inferred for this phase \citep{Wetzel+13}. 
The $\gtrsim$4\,Gyr needed to erode the hot halo matches typical satellite quenching timescales \citep[e.g.,][]{Wetzel+13,Peng+15,Pasquali+19,Oman+21,deIsidio+26}, and the star-formation depletion time estimated at infall ($\tau_{\rm dep}=M_{\rm cool~gas}/{\rm SFR}\approx5.5^{+0.2}_{-0.1}$\,Gyr; Appendix\,\ref{AppendixC}) is of the same order, indicating that once cooling is curtailed the cool gas is exhausted on essentially its consumption timescale. We therefore identify the progressive disconnection from the CGM as the dominant factor setting the long-term evolution of satellites rather than the abrupt destruction of the cold disk, with infall marking the onset of a gradual, several-Gyr starvation process.

\subsection{Metallicity enhancement and structural compaction as signatures of tidal stripping of the outer layers}
\label{Metal_enhancement}

The metallicity evolution of satellites offers an independent probe of environmental quenching. Our data reproduce the observable that motivates the starvation picture of \citet{Peng+15}: satellites are systematically more metal-rich than centrals at fixed stellar mass, with the offset growing with time since infall (Fig.\,\ref{fig:metallicity}). As shown in $\S$\ref{subsec:metallicity}, this metallicity is set primarily by the time a satellite has spent in the host rather than by its stellar mass, the broad scatter at $M_\star $$\,\lesssim\,$$ 10^{10.5}\,\rm M_\odot$ reflecting a spread in infall times. 
We also note that the same weak mass dependence and infall-time correlation appear in SDSS phase-space studies and in simulations of the inherited metallicity--environment relation \citep[e.g.,][]{Genel+16,Pasquali+19}.
The strong stellar-metallicity enhancement we find, together with its tight link to structural compaction, therefore points beyond pure starvation.
In IllustrisTNG, the enhanced metallicities, the structural compaction, and the suppression of the gas supply appear to emerge together as the consequence of the progressive tidal stripping of the loosely bound outer material of satellites, from the gaseous and stellar outskirts down to the dark-matter halo itself.

\begin{figure}[htbp!]
  \centering
\includegraphics[width=0.46\textwidth]
{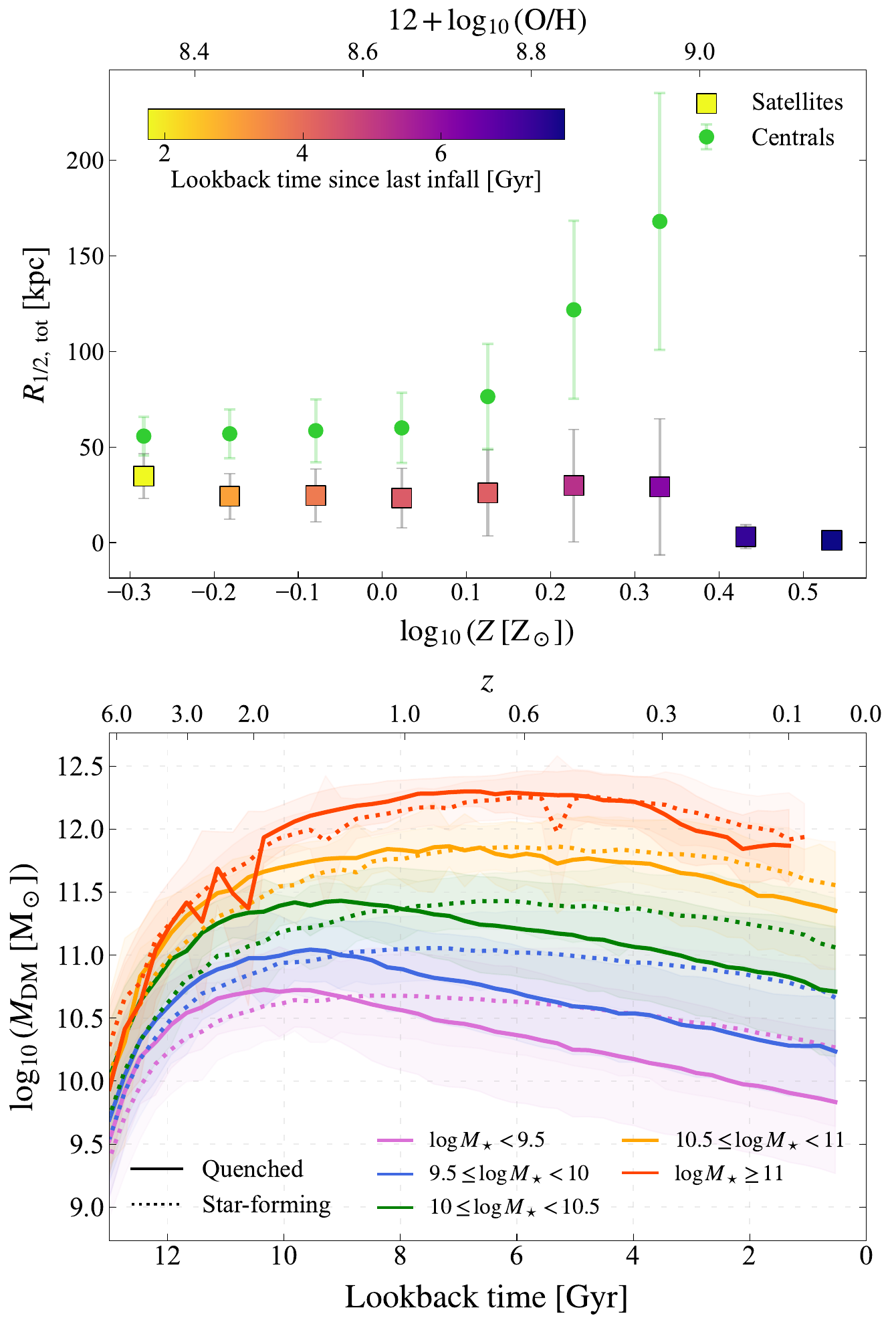}
    \caption{\textit{Upper~panel:} Size--metallicity relation of our sample as in Fig.\,\ref{fig:metallicity}, but using $R_{1/2,\,\rm tot}$, the radius enclosing half of the galaxy's total (baryonic $+$ dark matter) mass.
    Satellites are color-coded by the lookback time since the last infall. The contrast between the compact satellites and the more extended centrals is even more pronounced than for the stellar half-mass radius (Fig.\,\ref{fig:metallicity}c), indicating that the compaction extends to the dark-matter--dominated mass.
    \textit{Lower~panel:} Dark-matter mass evolution of satellites as a function of lookback time, split into five present-day ($z$$\,\sim\,$$0$) stellar-mass bins. Solid curves represent quenched satellites, whereas dotted curves show currently star-forming galaxies. 
    }
    \label{fig:DM}
\end{figure}

The clearest signature of this processing is structural. The most metal-rich satellites are also the most compact, in contrast to centrals, whose sizes increase with metallicity (stellar half-mass radius; Fig.\,\ref{fig:metallicity}c). Crucially, this is not a purely baryonic effect: when size is instead measured by the total, dark-matter–dominated half-mass radius, the same satellite–central contrast becomes even more pronounced (upper panel of Fig.\,\ref{fig:DM}). The compaction of the most metal-rich satellites is therefore more clearly visible in the dark matter than in the stars, indicating that the halo itself, and not only the baryonic body, is affected.
This is decisive for identifying the mechanism: ram pressure and other hydrodynamical processes act on the gas and are not expected to displace stars or dark matter, so the compaction of both the stellar and the dark-matter–dominated half-mass radii, together with the post-infall decline of the dark-matter mass (lower panel of Fig.\,\ref{fig:DM}), must be essentially gravitational. 
These structural signatures therefore strongly suggest that it is tidal stripping, not hydrodynamical gas stripping processes (e.g., ram pressure), that best explains the outer layer removal of satellites.

The assembly history of the dark matter confirms this. Rather than growing monotonically, we show in the lower panel of Fig.\,\ref{fig:DM} that the dark-matter mass of satellites rises until around the epoch of infall and then declines toward the present day, which is a fingerprint of tidal stripping within the host potential. The decline is strongest for the currently quenched and lowest-mass satellites, which lose the largest fraction of their halo, while more massive systems with deeper potential wells retain theirs. 

These results point to a mechanism operating beneath the observed starvation signature. As the outer dark matter is stripped, the satellite's potential well becomes shallower, reducing the binding energy of the surrounding gas so that both the hot reservoir and the loosely bound outer disk are more easily removed by ram pressure and tides, suppressing further accretion. The loss of dark matter therefore does not merely accompany gas depletion but actively facilitates it: the pronounced halo stripping of the quenched population is consistent with a chain in which the weakening of the potential promotes the removal of the very gas whose exhaustion ultimately quenches the galaxy.

The chemical outcome follows from the negative metallicity gradients of disks:
removing outer material raises the mean metallicity while reducing the physical size, producing the compact, metal-rich, gas-poor remnants that the oldest satellites resemble.
This also clarifies why stripping leaves only an indirect imprint on the mass--metallicity relation, leaving behind systems that are metal-rich for their present-day stellar mass rather than displacing them along it.

\subsection{Caveats}
The main limitation of this work concerns the spatially resolved gas kinematics in TNG50. As discussed in $\S$\ref{Sec3}, the gas velocity maps of MaNGIA mock galaxies are systematically more disturbed than those observed in MaNGA, making gas kinematic asymmetries unreliable diagnostics in the simulation. This mismatch likely reflects a combination of feedback prescriptions, unresolved ISM physics, and forward-modeling limitations. We therefore exclude resolved gas kinematics from our analysis and instead base our conclusions regarding environmental stripping on the evolution of the global gas reservoir.

A second caveat concerns the hot-CGM depletion timescales derived in $\S$\ref{hot_gas}. While the measured decline of the hot halo with time since infall is robust within TNG50, the absolute values depend on the adopted hydrodynamical model, feedback implementation, and our definition of hot gas ($T$$\,\geq\,$$ 10^{5}$\,K). These timescales should therefore be interpreted as characteristic model-dependent quantities rather than precise observational predictions.

More generally, CGM properties remain sensitive to feedback prescriptions, numerical resolution, and the definition of gravitationally bound gas. Consequently, our quantitative estimates of hot-gas depletion are specific to the IllustrisTNG framework, although the qualitative picture of progressive CGM removal after infall is expected to be robust.

Finally, the limited volume of TNG50 under-samples the richest cluster environments, making our conclusions most representative of group-scale systems (including high-mass groups of $M_\star$$\,\sim\,$$\rm 10^{14}\,M_\odot$). In addition, quantities such as infall time and hot gas mass are theoretical constructs derived directly from the simulation and do not have direct observational counterparts. Likewise, simulated metallicities are known to exhibit slight offsets relative to observations \cite[e.g.,][]{Sarmiento+23}; however, because our analysis relies primarily on differential comparisons between satellites and centrals within the same simulation, the impact of these offsets should be minimized.

\section{Summary and Conclusions}
\label{Sec5_conclusions}
In this work, we used the MaNGA-like mock sample \citep[MaNGIA;][]{Sarmiento+23}, built by forward-modeling TNG50 galaxies through the MaNGA instrumental pipeline, to disentangle the physical mechanisms responsible for satellite quenching, with particular attention to the role of the circumgalactic medium. From a parent sample of more than 10\,000 mock IFS data cubes we selected a subsample of $\sim$7\,300 spatially resolved galaxies (2\,800 satellites) and reconstructed their full accretion, hot- and cool-gas, dark-matter, structural, and chemical histories along their \textsc{Sublink} merger trees, using the time since infall as the physical axis along which quenching unfolds.
 
Bringing these results together, satellite quenching in IllustrisTNG emerges not as a single mechanism but as a coupled, self-reinforcing cycle acting over several gigayears. 
At its core lies starvation: after infall, the hot CGM---the long-term fuel reservoir---is progressively removed, disconnecting the satellite from the baryon cycle, so that once cooling is curtailed the residual cool gas is simply exhausted by ongoing star formation on its consumption timescale ($\tau_{\rm dep}$$\,\approx\,$$5.5^{+0.2}_{-0.1}$\,Gyr). 
This depletion does not occur in isolation. Tidal interactions within the host halo strip the subhalo's dark matter, weakening its potential well and lowering the binding energy of the surrounding gas, which makes the hot CGM progressively easier to remove. 
The same tidal processing preferentially strips the loosely bound outer layers, driving the compaction of both the stellar and the dark-matter–dominated components and increasing the satellites' mean stellar metallicity;
starvation reinforces this chemical enhancement, but the accompanying compaction, which starvation alone cannot produce, identifies tidal stripping as an important driver. 
Other hydrodynamical processes (e.g., ram-pressure and viscous stripping, stellar and AGN feedback) may still act episodically, but the long-term outcome is governed by the progressive exhaustion of the circumgalactic reservoir and the structural and chemical transformation that accompanies it, leaving behind a compact, metal-rich, gas-starved and quenched satellite whose ordered stellar kinematics remain mostly intact.
 
Our main conclusions are as follows:
\begin{enumerate}[label=\roman*.]
\item The vast majority of satellites retain regular, rotation-supported stellar kinematics throughout their quenching phase; disturbed velocity fields are confined to low-mass systems, with the most extreme asymmetries below $M_\star$$\,\sim\,$$10^{10}\,\mathrm{M_\odot}$. The dominant quenching pathways therefore leave the collisionless stellar body largely undisturbed \citep[e.g.,][\S\ref{symmetric}]{Cortese+19,deIsidio+26}.
 
\item We provide, for the first time, a clear time evolution of hot-CGM depletion after infall: satellites lose $\sim$50\% of their hot gas reservoir within $\sim$2\,Gyr---coinciding with the expected first pericentric passage---and $\sim$90\% within $4.2^{+0.6}_{-0.6}$\,Gyr, with the depletion increasing monotonically with residence time and essentially independent of stellar mass ($\S$\ref{hot_gas};\,Appendix\,\ref{AppendixB}).
 
\item The hot gas mass correlates strongly with SFR, establishing the hot CGM as the long-term fuel reservoir whose progressive exhaustion regulates star formation in satellites. Quenched centrals, by contrast, retain massive hot halos ($\log_{10}(M_{\rm hot~gas}/\rm M_\odot)$$\,\sim\,$$10$--$11$) despite their low SFR, pointing to distinct channels: environmental stripping and starvation in satellites versus maintenance-mode AGN feedback in centrals ($\S$\ref{hot_gas};\,Appendix\,\ref{AppendixD}).
 
\item Present-day quenched satellites were accreted significantly earlier than star-forming ones ($6.5^{+0.3}_{-0.3}$ vs.\ $4.3^{+0.3}_{-0.3}$\,Gyr ago), continuing to form stars for $\gtrsim$3\,Gyr after infall before declining sharply---a delayed-then-rapid pathway also traced by the two-phase decline of the cool gas ($\S$\ref{hot_gas},\,$\S$\ref{rapidthendelayed};\,Appendix\,\ref{AppendixC}).
 
\item Satellites lose little stellar mass but are tidally stripped of dark matter and of their metal-poor outer layers: the dark-matter mass declines after infall (most strongly in quenched, low-mass systems) and the compaction is even more pronounced in the total, dark-matter--dominated half-mass radius than in the stars (Fig.\,\ref{fig:DM}). 
Because ram pressure and other hydrodynamical processes act on the gas and are not expected to displace stars or dark matter, this structural transformation and the accompanying increase in stellar metallicity is best understood as a signature of \emph{tidal} stripping; in this scenario, the tidal loss of dark matter also shallows the potential well and facilitates removal of the hot CGM, rather than merely accompanying it (\S\ref{subsec:metallicity}, \S\ref{Metal_enhancement}).
\end{enumerate}

These conclusions describe the quenching pathways realized by the IllustrisTNG model and are expected to be most representative of group-scale environments; direct observational tests will require proxies for the bound hot-gas reservoir and satellite infall time.

\begin{acknowledgements}
    NdI gratefully acknowledges the IMPRS program and ESO for the support and funding of his PhD.
    PP acknowledges financial support from the European Research Council (ERC) under the European Union’s Horizon Europe research and innovation programme ERC CoG CLEVeR (Grant agreement No. 101045437). NdI also thanks Felix Heinze for his assistance with the infall time algorithm and Frank van den Bosch for the insightful discussions.
    
\end{acknowledgements}
  
\small
\bibliographystyle{aa} 
\bibliography{lib.bib}

@ARTICLE{Tacchella+22a,
       author = {{Tacchella}, Sandro and {Conroy}, Charlie and {Faber}, S.~M. and {Johnson}, Benjamin D. and {Leja}, Joel and {Barro}, Guillermo and {Cunningham}, Emily C. and {Deason}, Alis J. and {Guhathakurta}, Puragra and {Guo}, Yicheng and {Hernquist}, Lars and {Koo}, David C. and {McKinnon}, Kevin and {Rockosi}, Constance M. and {Speagle}, Joshua S. and {van Dokkum}, Pieter and {Yesuf}, Hassen M.},
        title = "{Fast, Slow, Early, Late: Quenching Massive Galaxies at z {\ensuremath{\sim}} 0.8}",
      journal = {\apj},
         year = 2022,
        month = feb,
       volume = {926},
       number = {2},
          eid = {134},
        pages = {134},
          doi = {10.3847/1538-4357/ac449b},
archivePrefix = {arXiv},
       eprint = {2102.12494},
 primaryClass = {astro-ph.GA},
       adsurl = {https://ui.adsabs.harvard.edu/abs/2022ApJ...926..134T}
}

@ARTICLE{Peeples+19,
       author = {{Peeples}, Molly S. and {Corlies}, Lauren and {Tumlinson}, Jason and {O'Shea}, Brian W. and {Lehner}, Nicolas and {O'Meara}, John M. and {Howk}, J. Christopher and {Earl}, Nicholas and {Smith}, Britton D. and {Wise}, John H. and {Hummels}, Cameron B.},
        title = "{Figuring Out Gas \& Galaxies in Enzo (FOGGIE). I. Resolving Simulated Circumgalactic Absorption at 2 {\ensuremath{\leq}} z {\ensuremath{\leq}} 2.5}",
      journal = {\apj},
         year = 2019,
        month = mar,
       volume = {873},
       number = {2},
          eid = {129},
        pages = {129},
          doi = {10.3847/1538-4357/ab0654},
archivePrefix = {arXiv},
       eprint = {1810.06566},
 primaryClass = {astro-ph.GA},
       adsurl = {https://ui.adsabs.harvard.edu/abs/2019ApJ...873..129P}
}

@ARTICLE{Bahe+13,
       author = {{Bah{\'e}}, Yannick M. and {McCarthy}, Ian G. and {Balogh}, Michael L. and {Font}, Andreea S.},
        title = "{Why does the environmental influence on group and cluster galaxies extend beyond the virial radius?}",
      journal = {\mnras},
         year = 2013,
        month = apr,
       volume = {430},
       number = {4},
        pages = {3017-3031},
          doi = {10.1093/mnras/stt109},
archivePrefix = {arXiv},
       eprint = {1210.8407},
 primaryClass = {astro-ph.CO},
       adsurl = {https://ui.adsabs.harvard.edu/abs/2013MNRAS.430.3017B}
}

@ARTICLE{Tonnesen+09,
       author = {{Tonnesen}, Stephanie and {Bryan}, Greg L.},
        title = "{Gas Stripping in Simulated Galaxies with a Multiphase Interstellar Medium}",
      journal = {\apj},
         year = 2009,
        month = apr,
       volume = {694},
       number = {2},
        pages = {789-804},
          doi = {10.1088/0004-637X/694/2/789},
archivePrefix = {arXiv},
       eprint = {0901.2115},
 primaryClass = {astro-ph.GA},
       adsurl = {https://ui.adsabs.harvard.edu/abs/2009ApJ...694..789T}
}

@ARTICLE{Barrera-Ballesteros+15,
       author = {{Barrera-Ballesteros}, J.~K. and {Garc{\'\i}a-Lorenzo}, B. and {Falc{\'o}n-Barroso}, J. and {van de Ven}, G. and {Lyubenova}, M. and {Wild}, V. and {M{\'e}ndez-Abreu}, J. and {S{\'a}nchez}, S.~F. and {Marquez}, I. and {Masegosa}, J. and {Monreal-Ibero}, A. and {Ziegler}, B. and {del Olmo}, A. and {Verdes-Montenegro}, L. and {Garc{\'\i}a-Benito}, R. and {Husemann}, B. and {Mast}, D. and {Kehrig}, C. and {Iglesias-Paramo}, J. and {Marino}, R.~A. and {Aguerri}, J.~A.~L. and {Walcher}, C.~J. and {V{\'\i}lchez}, J.~M. and {Bomans}, D.~J. and {Cortijo-Ferrero}, C. and {Gonz{\'a}lez Delgado}, R.~M. and {Bland-Hawthorn}, J. and {McIntosh}, D.~H. and {Bekerait{\.{e}}}, S.},
        title = "{Tracing kinematic (mis)alignments in CALIFA merging galaxies. Stellar and ionized gas kinematic orientations at every merger stage}",
      journal = {\aap},
         year = 2015,
        month = oct,
       volume = {582},
          eid = {A21},
        pages = {A21},
          doi = {10.1051/0004-6361/201424935},
archivePrefix = {arXiv},
       eprint = {1506.03819},
 primaryClass = {astro-ph.GA},
       adsurl = {https://ui.adsabs.harvard.edu/abs/2015A&A...582A..21B}
}

@ARTICLE{Weinberger+17,
       author = {{Weinberger}, Rainer and {Springel}, Volker and {Hernquist}, Lars and {Pillepich}, Annalisa and {Marinacci}, Federico and {Pakmor}, R{\"u}diger and {Nelson}, Dylan and {Genel}, Shy and {Vogelsberger}, Mark and {Naiman}, Jill and {Torrey}, Paul},
        title = "{Simulating galaxy formation with black hole driven thermal and kinetic feedback}",
      journal = {\mnras},
         year = 2017,
        month = mar,
       volume = {465},
       number = {3},
        pages = {3291-3308},
          doi = {10.1093/mnras/stw2944},
archivePrefix = {arXiv},
       eprint = {1607.03486},
 primaryClass = {astro-ph.GA},
       adsurl = {https://ui.adsabs.harvard.edu/abs/2017MNRAS.465.3291W}
}

@ARTICLE{Pillepich+18,
       author = {{Pillepich}, Annalisa and {Nelson}, Dylan and {Hernquist}, Lars and {Springel}, Volker and {Pakmor}, R{\"u}diger and {Torrey}, Paul and {Weinberger}, Rainer and {Genel}, Shy and {Naiman}, Jill P. and {Marinacci}, Federico and {Vogelsberger}, Mark},
        title = "{First results from the IllustrisTNG simulations: the stellar mass content of groups and clusters of galaxies}",
      journal = {\mnras},
         year = 2018,
        month = mar,
       volume = {475},
       number = {1},
        pages = {648-675},
          doi = {10.1093/mnras/stx3112},
archivePrefix = {arXiv},
       eprint = {1707.03406},
 primaryClass = {astro-ph.GA},
       adsurl = {https://ui.adsabs.harvard.edu/abs/2018MNRAS.475..648P}
}

@ARTICLE{Bloom+18,
       author = {{Bloom}, J.~V. and {Croom}, S.~M. and {Bryant}, J.~J. and {Schaefer}, A.~L. and {Bland-Hawthorn}, J. and {Brough}, S. and {Callingham}, J. and {Cortese}, L. and {Federrath}, C. and {Scott}, N. and {van de Sande}, J. and {D'Eugenio}, F. and {Sweet}, S. and {Tonini}, C. and {Allen}, J.~T. and {Goodwin}, M. and {Green}, A.~W. and {Konstantopoulos}, I.~S. and {Lawrence}, J. and {Lorente}, N. and {Medling}, A.~M. and {Owers}, M.~S. and {Richards}, S.~N. and {Sharp}, R.},
        title = "{The SAMI Galaxy Survey: gas content and interaction as the drivers of kinematic asymmetry}",
      journal = {\mnras},
         year = 2018,
        month = may,
       volume = {476},
       number = {2},
        pages = {2339-2351},
          doi = {10.1093/mnras/sty273},
archivePrefix = {arXiv},
       eprint = {1801.06628},
 primaryClass = {astro-ph.GA},
       adsurl = {https://ui.adsabs.harvard.edu/abs/2018MNRAS.476.2339B}
}

@ARTICLE{Shapiro+08,
       author = {{Shapiro}, Kristen L. and {Genzel}, Reinhard and {F{\"o}rster Schreiber}, Natascha M. and {Tacconi}, Linda J. and {Bouch{\'e}}, Nicolas and {Cresci}, Giovanni and {Davies}, Richard and {Eisenhauer}, Frank and {Johansson}, Peter H. and {Krajnovi{\'c}}, Davor and {Lutz}, Dieter and {Naab}, Thorsten and {Arimoto}, Nobuo and {Arribas}, Santiago and {Cimatti}, Andrea and {Colina}, Luis and {Daddi}, Emanuele and {Daigle}, Olivier and {Erb}, Dawn and {Hernandez}, Olivier and {Kong}, Xu and {Mignoli}, Marco and {Onodera}, Masato and {Renzini}, Alvio and {Shapley}, Alice and {Steidel}, Charles},
        title = "{Kinemetry of SINS High-Redshift Star-Forming Galaxies: Distinguishing Rotating Disks from Major Mergers}",
      journal = {\apj},
         year = 2008,
        month = jul,
       volume = {682},
       number = {1},
        pages = {231-251},
          doi = {10.1086/587133},
archivePrefix = {arXiv},
       eprint = {0802.0879},
 primaryClass = {astro-ph},
       adsurl = {https://ui.adsabs.harvard.edu/abs/2008ApJ...682..231S}
}

@ARTICLE{Bloom+17,
       author = {{Bloom}, J.~V. and {Fogarty}, L.~M.~R. and {Croom}, S.~M. and {Schaefer}, A. and {Bryant}, J.~J. and {Cortese}, L. and {Richards}, S. and {Bland-Hawthorn}, J. and {Ho}, I. -T. and {Scott}, N. and {Goldstein}, G. and {Medling}, A. and {Brough}, S. and {Sweet}, S.~M. and {Cecil}, G. and {L{\'o}pez-S{\'a}nchez}, A. and {Glazebrook}, K. and {Parker}, Q. and {Allen}, J.~T. and {Goodwin}, M. and {Green}, A.~W. and {Konstantopoulos}, I.~S. and {Lawrence}, J.~S. and {Lorente}, N. and {Owers}, M.~S. and {Sharp}, R.},
        title = "{The SAMI Galaxy Survey: asymmetry in gas kinematics and its links to stellar mass and star formation}",
      journal = {\mnras},
         year = 2017,
        month = feb,
       volume = {465},
       number = {1},
        pages = {123-148},
          doi = {10.1093/mnras/stw2605},
archivePrefix = {arXiv},
       eprint = {1610.02773},
 primaryClass = {astro-ph.GA},
       adsurl = {https://ui.adsabs.harvard.edu/abs/2017MNRAS.465..123B}
}

@ARTICLE{McClymont+24,
       author = {{McClymont}, William and {Tacchella}, Sandro and {Smith}, Aaron and {Kannan}, Rahul and {Maiolino}, Roberto and {Belfiore}, Francesco and {Hernquist}, Lars and {Li}, Hui and {Vogelsberger}, Mark},
        title = "{The nature of diffuse ionized gas in star-forming galaxies}",
      journal = {\mnras},
         year = 2024,
        month = aug,
       volume = {532},
       number = {2},
        pages = {2016-2031},
          doi = {10.1093/mnras/stae1587},
archivePrefix = {arXiv},
       eprint = {2403.03243},
 primaryClass = {astro-ph.GA},
       adsurl = {https://ui.adsabs.harvard.edu/abs/2024MNRAS.532.2016M}
}

@ARTICLE{Pasquali+12,
       author = {{Pasquali}, Anna and {Gallazzi}, Anna and {van den Bosch}, Frank C.},
        title = "{The gas-phase metallicity of central and satellite galaxies in the Sloan Digital Sky Survey}",
      journal = {\mnras},
         year = 2012,
        month = sep,
       volume = {425},
       number = {1},
        pages = {273-286},
          doi = {10.1111/j.1365-2966.2012.21454.x},
archivePrefix = {arXiv},
       eprint = {1206.3458},
 primaryClass = {astro-ph.CO},
       adsurl = {https://ui.adsabs.harvard.edu/abs/2012MNRAS.425..273P}
}

@ARTICLE{Tacchella+22c,
       author = {{Tacchella}, Sandro and {Smith}, Aaron and {Kannan}, Rahul and {Marinacci}, Federico and {Hernquist}, Lars and {Vogelsberger}, Mark and {Torrey}, Paul and {Sales}, Laura and {Li}, Hui},
        title = "{H {\ensuremath{\alpha}} emission in local galaxies: star formation, time variability, and the diffuse ionized gas}",
      journal = {\mnras},
         year = 2022,
        month = jun,
       volume = {513},
       number = {2},
        pages = {2904-2929},
          doi = {10.1093/mnras/stac818},
archivePrefix = {arXiv},
       eprint = {2112.00027},
 primaryClass = {astro-ph.GA},
       adsurl = {https://ui.adsabs.harvard.edu/abs/2022MNRAS.513.2904T}
}

@ARTICLE{PlanckCollaboration+16,
       author = {{Planck Collaboration XIII} and {Ade}, P.~A.~R. and {Aghanim}, N. and {Arnaud}, M. and {Ashdown}, M. and {Aumont}, J. and {Baccigalupi}, C. and {Banday}, A.~J. and {Barreiro}, R.~B. and {Bartlett}, J.~G. and {Bartolo}, N. and {Battaner}, E. and {Battye}, R. and {Benabed}, K. and {Beno{\^\i}t}, A. and {Benoit-L{\'e}vy}, A. and {Bernard}, J. -P. and {Bersanelli}, M. and {Bielewicz}, P. and {Bock}, J.~J. and {Bonaldi}, A. and {Bonavera}, L. and {Bond}, J.~R. and {Borrill}, J. and {Bouchet}, F.~R. and {Boulanger}, F. and {Bucher}, M. and {Burigana}, C. and {Butler}, R.~C. and {Calabrese}, E. and {Cardoso}, J. -F. and {Catalano}, A. and {Challinor}, A. and {Chamballu}, A. and {Chary}, R. -R. and {Chiang}, H.~C. and {Chluba}, J. and {Christensen}, P.~R. and {Church}, S. and {Clements}, D.~L. and {Colombi}, S. and {Colombo}, L.~P.~L. and {Combet}, C. and {Coulais}, A. and {Crill}, B.~P. and {Curto}, A. and {Cuttaia}, F. and {Danese}, L. and {Davies}, R.~D. and {Davis}, R.~J. and {de Bernardis}, P. and {de Rosa}, A. and {de Zotti}, G. and {Delabrouille}, J. and {D{\'e}sert}, F. -X. and {Di Valentino}, E. and {Dickinson}, C. and {Diego}, J.~M. and {Dolag}, K. and {Dole}, H. and {Donzelli}, S. and {Dor{\'e}}, O. and {Douspis}, M. and {Ducout}, A. and {Dunkley}, J. and {Dupac}, X. and {Efstathiou}, G. and {Elsner}, F. and {En{\ss}lin}, T.~A. and {Eriksen}, H.~K. and {Farhang}, M. and {Fergusson}, J. and {Finelli}, F. and {Forni}, O. and {Frailis}, M. and {Fraisse}, A.~A. and {Franceschi}, E. and {Frejsel}, A. and {Galeotta}, S. and {Galli}, S. and {Ganga}, K. and {Gauthier}, C. and {Gerbino}, M. and {Ghosh}, T. and {Giard}, M. and {Giraud-H{\'e}raud}, Y. and {Giusarma}, E. and {Gjerl{\o}w}, E. and {Gonz{\'a}lez-Nuevo}, J. and {G{\'o}rski}, K.~M. and {Gratton}, S. and {Gregorio}, A. and {Gruppuso}, A. and {Gudmundsson}, J.~E. and {Hamann}, J. and {Hansen}, F.~K. and {Hanson}, D. and {Harrison}, D.~L. and {Helou}, G. and {Henrot-Versill{\'e}}, S. and {Hern{\'a}ndez-Monteagudo}, C. and {Herranz}, D. and {Hildebrandt}, S.~R. and {Hivon}, E. and {Hobson}, M. and {Holmes}, W.~A. and {Hornstrup}, A. and {Hovest}, W. and {Huang}, Z. and {Huffenberger}, K.~M. and {Hurier}, G. and {Jaffe}, A.~H. and {Jaffe}, T.~R. and {Jones}, W.~C. and {Juvela}, M. and {Keih{\"a}nen}, E. and {Keskitalo}, R. and {Kisner}, T.~S. and {Kneissl}, R. and {Knoche}, J. and {Knox}, L. and {Kunz}, M. and {Kurki-Suonio}, H. and {Lagache}, G. and {L{\"a}hteenm{\"a}ki}, A. and {Lamarre}, J. -M. and {Lasenby}, A. and {Lattanzi}, M. and {Lawrence}, C.~R. and {Leahy}, J.~P. and {Leonardi}, R. and {Lesgourgues}, J. and {Levrier}, F. and {Lewis}, A. and {Liguori}, M. and {Lilje}, P.~B. and {Linden-V{\o}rnle}, M. and {L{\'o}pez-Caniego}, M. and {Lubin}, P.~M. and {Mac{\'\i}as-P{\'e}rez}, J.~F. and {Maggio}, G. and {Maino}, D. and {Mandolesi}, N. and {Mangilli}, A. and {Marchini}, A. and {Maris}, M. and {Martin}, P.~G. and {Martinelli}, M. and {Mart{\'\i}nez-Gonz{\'a}lez}, E. and {Masi}, S. and {Matarrese}, S. and {McGehee}, P. and {Meinhold}, P.~R. and {Melchiorri}, A. and {Melin}, J. -B. and {Mendes}, L. and {Mennella}, A. and {Migliaccio}, M. and {Millea}, M. and {Mitra}, S. and {Miville-Desch{\^e}nes}, M. -A. and {Moneti}, A. and {Montier}, L. and {Morgante}, G. and {Mortlock}, D. and {Moss}, A. and {Munshi}, D. and {Murphy}, J.~A. and {Naselsky}, P. and {Nati}, F. and {Natoli}, P. and {Netterfield}, C.~B. and {N{\o}rgaard-Nielsen}, H.~U. and {Noviello}, F. and {Novikov}, D. and {Novikov}, I. and {Oxborrow}, C.~A. and {Paci}, F. and {Pagano}, L. and {Pajot}, F. and {Paladini}, R. and {Paoletti}, D. and {Partridge}, B. and {Pasian}, F. and {Patanchon}, G. and {Pearson}, T.~J. and {Perdereau}, O. and {Perotto}, L. and {Perrotta}, F. and {Pettorino}, V. and {Piacentini}, F. and {Piat}, M. and {Pierpaoli}, E. and {Pietrobon}, D. and {Plaszczynski}, S. and {Pointecouteau}, E. and {Polenta}, G. and {Popa}, L. and {Pratt}, G.~W. and {Pr{\'e}zeau}, G.},
        title = "{Planck 2015 results. XIII. Cosmological parameters}",
      journal = {\aap},
         year = 2016,
        month = sep,
       volume = {594},
          eid = {A13},
        pages = {A13},
          doi = {10.1051/0004-6361/201525830},
archivePrefix = {arXiv},
       eprint = {1502.01589},
 primaryClass = {astro-ph.CO},
       adsurl = {https://ui.adsabs.harvard.edu/abs/2016A&A...594A..13P}
}

@ARTICLE{deIsidio+26,
       author = {{de Is{\'\i}dio}, Natan and {Popesso}, P. and {Bah{\'e}}, Y. and {Vulcani}, B. and {Toptun}, V. and {Marini}, I. and {Poggianti}, B. and {Biffi}, V. and {Belfiore}, F. and {Lagos}, C. and {Dolag}, K. and {Mazengo}, D.},
        title = "{The kinematic imprinting of environmental quenching in $z<0.2$ galaxies}",
      journal = {arXiv e-prints},
         year = 2026,
        month = mar,
          eid = {arXiv:2603.03432},
        pages = {arXiv:2603.03432},
          doi = {10.48550/arXiv.2603.03432},
archivePrefix = {arXiv},
       eprint = {2603.03432},
 primaryClass = {astro-ph.GA},
       adsurl = {https://ui.adsabs.harvard.edu/abs/2026arXiv260303432D}
}

@ARTICLE{Pillepich+19,
       author = {{Pillepich}, Annalisa and {Nelson}, Dylan and {Springel}, Volker and {Pakmor}, R{\"u}diger and {Torrey}, Paul and {Weinberger}, Rainer and {Vogelsberger}, Mark and {Marinacci}, Federico and {Genel}, Shy and {van der Wel}, Arjen and {Hernquist}, Lars},
        title = "{First results from the TNG50 simulation: the evolution of stellar and gaseous discs across cosmic time}",
      journal = {\mnras},
         year = 2019,
        month = dec,
       volume = {490},
       number = {3},
        pages = {3196-3233},
          doi = {10.1093/mnras/stz2338},
archivePrefix = {arXiv},
       eprint = {1902.05553},
 primaryClass = {astro-ph.GA},
       adsurl = {https://ui.adsabs.harvard.edu/abs/2019MNRAS.490.3196P}
}

@ARTICLE{Nelson+19,
       author = {{Nelson}, Dylan and {Pillepich}, Annalisa and {Springel}, Volker and {Pakmor}, R{\"u}diger and {Weinberger}, Rainer and {Genel}, Shy and {Torrey}, Paul and {Vogelsberger}, Mark and {Marinacci}, Federico and {Hernquist}, Lars},
        title = "{First results from the TNG50 simulation: galactic outflows driven by supernovae and black hole feedback}",
      journal = {\mnras},
         year = 2019,
        month = dec,
       volume = {490},
       number = {3},
        pages = {3234-3261},
          doi = {10.1093/mnras/stz2306},
archivePrefix = {arXiv},
       eprint = {1902.05554},
 primaryClass = {astro-ph.GA},
       adsurl = {https://ui.adsabs.harvard.edu/abs/2019MNRAS.490.3234N}
}

@ARTICLE{Sarmiento+23,
       author = {{Sarmiento}, Regina and {Huertas-Company}, Marc and {Knapen}, Johan H. and {Ibarra-Medel}, H{\'e}ctor and {Pillepich}, Annalisa and {S{\'a}nchez}, Sebasti{\'a}n F. and {Boecker}, Alina},
        title = "{MaNGIA: 10 000 mock galaxies for stellar population analysis}",
      journal = {\aap},
         year = 2023,
        month = may,
       volume = {673},
          eid = {A23},
        pages = {A23},
          doi = {10.1051/0004-6361/202245509},
archivePrefix = {arXiv},
       eprint = {2211.11790},
 primaryClass = {astro-ph.GA},
       adsurl = {https://ui.adsabs.harvard.edu/abs/2023A&A...673A..23S}
}

@ARTICLE{Rodriguez-Gomez+15,
       author = {{Rodriguez-Gomez}, Vicente and {Genel}, Shy and {Vogelsberger}, Mark and {Sijacki}, Debora and {Pillepich}, Annalisa and {Sales}, Laura V. and {Torrey}, Paul and {Snyder}, Greg and {Nelson}, Dylan and {Springel}, Volker and {Ma}, Chung-Pei and {Hernquist}, Lars},
        title = "{The merger rate of galaxies in the Illustris simulation: a comparison with observations and semi-empirical models}",
      journal = {\mnras},
         year = 2015,
        month = may,
       volume = {449},
       number = {1},
        pages = {49-64},
          doi = {10.1093/mnras/stv264},
archivePrefix = {arXiv},
       eprint = {1502.01339},
 primaryClass = {astro-ph.GA},
       adsurl = {https://ui.adsabs.harvard.edu/abs/2015MNRAS.449...49R}
}

@ARTICLE{Dolag+09,
       author = {{Dolag}, K. and {Borgani}, S. and {Murante}, G. and {Springel}, V.},
        title = "{Substructures in hydrodynamical cluster simulations}",
      journal = {\mnras},
         year = 2009,
        month = oct,
       volume = {399},
       number = {2},
        pages = {497-514},
          doi = {10.1111/j.1365-2966.2009.15034.x},
archivePrefix = {arXiv},
       eprint = {0808.3401},
 primaryClass = {astro-ph},
       adsurl = {https://ui.adsabs.harvard.edu/abs/2009MNRAS.399..497D}
}

@ARTICLE{Springel+01,
       author = {{Springel}, Volker and {White}, Simon D.~M. and {Tormen}, Giuseppe and {Kauffmann}, Guinevere},
        title = "{Populating a cluster of galaxies - I. Results at z=0}",
      journal = {\mnras},
         year = 2001,
        month = dec,
       volume = {328},
       number = {3},
        pages = {726-750},
          doi = {10.1046/j.1365-8711.2001.04912.x},
archivePrefix = {arXiv},
       eprint = {astro-ph/0012055},
 primaryClass = {astro-ph},
       adsurl = {https://ui.adsabs.harvard.edu/abs/2001MNRAS.328..726S}
}

@ARTICLE{Davis+85,
       author = {{Davis}, M. and {Efstathiou}, G. and {Frenk}, C.~S. and {White}, S.~D.~M.},
        title = "{The evolution of large-scale structure in a universe dominated by cold dark matter}",
      journal = {\apj},
         year = 1985,
        month = may,
       volume = {292},
        pages = {371-394},
          doi = {10.1086/163168},
       adsurl = {https://ui.adsabs.harvard.edu/abs/1985ApJ...292..371D}
}

@ARTICLE{Hidalgo-Pineda+26,
       author = {{Hidalgo-Pineda}, Fernando and {Gronke}, Max and {Grete}, Philipp},
        title = "{The launching of galactic winds from a multiphase ISM}",
      journal = {\mnras},
         year = 2026,
        month = may,
       volume = {548},
       number = {1},
          eid = {stag539},
        pages = {stag539},
          doi = {10.1093/mnras/stag539},
archivePrefix = {arXiv},
       eprint = {2510.14829},
 primaryClass = {astro-ph.GA},
       adsurl = {https://ui.adsabs.harvard.edu/abs/2026MNRAS.548ag539H}
}

@ARTICLE{Hayward+17,
       author = {{Hayward}, Christopher C. and {Hopkins}, Philip F.},
        title = "{How stellar feedback simultaneously regulates star formation and drives outflows}",
      journal = {\mnras},
         year = 2017,
        month = feb,
       volume = {465},
       number = {2},
        pages = {1682-1698},
          doi = {10.1093/mnras/stw2888},
archivePrefix = {arXiv},
       eprint = {1510.05650},
 primaryClass = {astro-ph.GA},
       adsurl = {https://ui.adsabs.harvard.edu/abs/2017MNRAS.465.1682H}
}

@ARTICLE{Heinze+24,
       author = {{Heinze}, Felix M. and {Despali}, Giulia and {Klessen}, Ralf S.},
        title = "{Not all subhaloes are created equal: modelling the diversity of subhalo density profiles in TNG50}",
      journal = {\mnras},
         year = 2024,
        month = feb,
       volume = {527},
       number = {4},
        pages = {11996-12015},
          doi = {10.1093/mnras/stad3894},
archivePrefix = {arXiv},
       eprint = {2311.13639},
 primaryClass = {astro-ph.GA},
       adsurl = {https://ui.adsabs.harvard.edu/abs/2024MNRAS.52711996H}
}

@ARTICLE{Asplund+09,
       author = {{Asplund}, Martin and {Grevesse}, Nicolas and {Sauval}, A. Jacques and {Scott}, Pat},
        title = "{The Chemical Composition of the Sun}",
      journal = {\araa},
         year = 2009,
        month = sep,
       volume = {47},
       number = {1},
        pages = {481-522},
          doi = {10.1146/annurev.astro.46.060407.145222},
archivePrefix = {arXiv},
       eprint = {0909.0948},
 primaryClass = {astro-ph.SR},
       adsurl = {https://ui.adsabs.harvard.edu/abs/2009ARA&A..47..481A}
}

@ARTICLE{deIsidio+24,
       author = {{de Is{\'\i}dio}, Natanael G. and {Men{\'e}ndez-Delmestre}, K. and {Gon{\c{c}}alves}, T.~S. and {Grossi}, M. and {Rodrigues}, D.~C. and {Garavito-Camargo}, N. and {Araujo-Carvalho}, A. and {Beaklini}, P.~P.~B. and {Cavalcante-Coelho}, Y. and {Cortesi}, A. and {Quiroga-Nu{\~n}ez}, L.~H. and {Randriamampandry}, T.},
        title = "{Dark Matter Distribution in Milky Way analog Galaxies}",
      journal = {\apj},
         year = 2024,
        month = aug,
       volume = {971},
       number = {1},
          eid = {69},
        pages = {69},
          doi = {10.3847/1538-4357/ad53c8},
archivePrefix = {arXiv},
       eprint = {2310.13839},
 primaryClass = {astro-ph.GA},
       adsurl = {https://ui.adsabs.harvard.edu/abs/2024ApJ...971...69D}
}

@ARTICLE{MaNGA+15,
       author = {{Bundy}, Kevin and {Bershady}, Matthew A. and {Law}, David R. and {Yan}, Renbin and {Drory}, Niv and {MacDonald}, Nicholas and {Wake}, David A. and {Cherinka}, Brian and {S{\'a}nchez-Gallego}, Jos{\'e} R. and {Weijmans}, Anne-Marie and {Thomas}, Daniel and {Tremonti}, Christy and {Masters}, Karen and {Coccato}, Lodovico and {Diamond-Stanic}, Aleksandar M. and {Arag{\'o}n-Salamanca}, Alfonso and {Avila-Reese}, Vladimir and {Badenes}, Carles and {Falc{\'o}n-Barroso}, J{\'e}sus and {Belfiore}, Francesco and {Bizyaev}, Dmitry and {Blanc}, Guillermo A. and {Bland-Hawthorn}, Joss and {Blanton}, Michael R. and {Brownstein}, Joel R. and {Byler}, Nell and {Cappellari}, Michele and {Conroy}, Charlie and {Dutton}, Aaron A. and {Emsellem}, Eric and {Etherington}, James and {Frinchaboy}, Peter M. and {Fu}, Hai and {Gunn}, James E. and {Harding}, Paul and {Johnston}, Evelyn J. and {Kauffmann}, Guinevere and {Kinemuchi}, Karen and {Klaene}, Mark A. and {Knapen}, Johan H. and {Leauthaud}, Alexie and {Li}, Cheng and {Lin}, Lihwai and {Maiolino}, Roberto and {Malanushenko}, Viktor and {Malanushenko}, Elena and {Mao}, Shude and {Maraston}, Claudia and {McDermid}, Richard M. and {Merrifield}, Michael R. and {Nichol}, Robert C. and {Oravetz}, Daniel and {Pan}, Kaike and {Parejko}, John K. and {Sanchez}, Sebastian F. and {Schlegel}, David and {Simmons}, Audrey and {Steele}, Oliver and {Steinmetz}, Matthias and {Thanjavur}, Karun and {Thompson}, Benjamin A. and {Tinker}, Jeremy L. and {van den Bosch}, Remco C.~E. and {Westfall}, Kyle B. and {Wilkinson}, David and {Wright}, Shelley and {Xiao}, Ting and {Zhang}, Kai},
        title = "{Overview of the SDSS-IV MaNGA Survey: Mapping nearby Galaxies at Apache Point Observatory}",
      journal = {\apj},
         year = 2015,
        month = jan,
       volume = {798},
       number = {1},
          eid = {7},
        pages = {7},
          doi = {10.1088/0004-637X/798/1/7},
archivePrefix = {arXiv},
       eprint = {1412.1482},
 primaryClass = {astro-ph.GA},
       adsurl = {https://ui.adsabs.harvard.edu/abs/2015ApJ...798....7B}
}

@ARTICLE{vandenBosch+08,
       author = {{van den Bosch}, Frank C. and {Aquino}, Daniel and {Yang}, Xiaohu and {Mo}, H.~J. and {Pasquali}, Anna and {McIntosh}, Daniel H. and {Weinmann}, Simone M. and {Kang}, Xi},
        title = "{The importance of satellite quenching for the build-up of the red sequence of present-day galaxies}",
      journal = {\mnras},
         year = 2008,
        month = jun,
       volume = {387},
       number = {1},
        pages = {79-91},
          doi = {10.1111/j.1365-2966.2008.13230.x},
archivePrefix = {arXiv},
       eprint = {0710.3164},
 primaryClass = {astro-ph},
       adsurl = {https://ui.adsabs.harvard.edu/abs/2008MNRAS.387...79V}
}

@ARTICLE{MaNGA+22,
       author = {{Abdurro'uf} and {Accetta}, Katherine and {Aerts}, Conny and {Silva Aguirre}, V{\'\i}ctor and {Ahumada}, Romina and {Ajgaonkar}, Nikhil and {Filiz Ak}, N. and {Alam}, Shadab and {Allende Prieto}, Carlos and {Almeida}, Andr{\'e}s and {Anders}, Friedrich and {Anderson}, Scott F. and {Andrews}, Brett H. and {Anguiano}, Borja and {Aquino-Ort{\'\i}z}, Erik and {Arag{\'o}n-Salamanca}, Alfonso and {Argudo-Fern{\'a}ndez}, Maria and {Ata}, Metin and {Aubert}, Marie and {Avila-Reese}, Vladimir and {Badenes}, Carles and {Barb{\'a}}, Rodolfo H. and {Barger}, Kat and {Barrera-Ballesteros}, Jorge K. and {Beaton}, Rachael L. and {Beers}, Timothy C. and {Belfiore}, Francesco and {Bender}, Chad F. and {Bernardi}, Mariangela and {Bershady}, Matthew A. and {Beutler}, Florian and {Bidin}, Christian Moni and {Bird}, Jonathan C. and {Bizyaev}, Dmitry and {Blanc}, Guillermo A. and {Blanton}, Michael R. and {Boardman}, Nicholas Fraser and {Bolton}, Adam S. and {Boquien}, M{\'e}d{\'e}ric and {Borissova}, Jura and {Bovy}, Jo and {Brandt}, W.~N. and {Brown}, Jordan and {Brownstein}, Joel R. and {Brusa}, Marcella and {Buchner}, Johannes and {Bundy}, Kevin and {Burchett}, Joseph N. and {Bureau}, Martin and {Burgasser}, Adam and {Cabang}, Tuesday K. and {Campbell}, Stephanie and {Cappellari}, Michele and {Carlberg}, Joleen K. and {Wanderley}, F{\'a}bio Carneiro and {Carrera}, Ricardo and {Cash}, Jennifer and {Chen}, Yan-Ping and {Chen}, Wei-Huai and {Cherinka}, Brian and {Chiappini}, Cristina and {Choi}, Peter Doohyun and {Chojnowski}, S. Drew and {Chung}, Haeun and {Clerc}, Nicolas and {Cohen}, Roger E. and {Comerford}, Julia M. and {Comparat}, Johan and {da Costa}, Luiz and {Covey}, Kevin and {Crane}, Jeffrey D. and {Cruz-Gonzalez}, Irene and {Culhane}, Connor and {Cunha}, Katia and {Dai}, Y. Sophia and {Damke}, Guillermo and {Darling}, Jeremy and {Davidson}, Jr., James W. and {Davies}, Roger and {Dawson}, Kyle and {De Lee}, Nathan and {Diamond-Stanic}, Aleksandar M. and {Cano-D{\'\i}az}, Mariana and {S{\'a}nchez}, Helena Dom{\'\i}nguez and {Donor}, John and {Duckworth}, Chris and {Dwelly}, Tom and {Eisenstein}, Daniel J. and {Elsworth}, Yvonne P. and {Emsellem}, Eric and {Eracleous}, Mike and {Escoffier}, Stephanie and {Fan}, Xiaohui and {Farr}, Emily and {Feng}, Shuai and {Fern{\'a}ndez-Trincado}, Jos{\'e} G. and {Feuillet}, Diane and {Filipp}, Andreas and {Fillingham}, Sean P. and {Frinchaboy}, Peter M. and {Fromenteau}, Sebastien and {Galbany}, Llu{\'\i}s and {Garc{\'\i}a}, Rafael A. and {Garc{\'\i}a-Hern{\'a}ndez}, D.~A. and {Ge}, Junqiang and {Geisler}, Doug and {Gelfand}, Joseph and {G{\'e}ron}, Tobias and {Gibson}, Benjamin J. and {Goddy}, Julian and {Godoy-Rivera}, Diego and {Grabowski}, Kathleen and {Green}, Paul J. and {Greener}, Michael and {Grier}, Catherine J. and {Griffith}, Emily and {Guo}, Hong and {Guy}, Julien and {Hadjara}, Massinissa and {Harding}, Paul and {Hasselquist}, Sten and {Hayes}, Christian R. and {Hearty}, Fred and {Hern{\'a}ndez}, Jes{\'u}s and {Hill}, Lewis and {Hogg}, David W. and {Holtzman}, Jon A. and {Horta}, Danny and {Hsieh}, Bau-Ching and {Hsu}, Chin-Hao and {Hsu}, Yun-Hsin and {Huber}, Daniel and {Huertas-Company}, Marc and {Hutchinson}, Brian and {Hwang}, Ho Seong and {Ibarra-Medel}, H{\'e}ctor J. and {Chitham}, Jacob Ider and {Ilha}, Gabriele S. and {Imig}, Julie and {Jaekle}, Will and {Jayasinghe}, Tharindu and {Ji}, Xihan and {Johnson}, Jennifer A. and {Jones}, Amy and {J{\"o}nsson}, Henrik and {Katkov}, Ivan and {Khalatyan}, Dr., Arman and {Kinemuchi}, Karen and {Kisku}, Shobhit and {Knapen}, Johan H. and {Kneib}, Jean-Paul and {Kollmeier}, Juna A. and {Kong}, Miranda and {Kounkel}, Marina and {Kreckel}, Kathryn and {Krishnarao}, Dhanesh and {Lacerna}, Ivan and {Lane}, Richard R. and {Langgin}, Rachel and {Lavender}, Ramon and {Law}, David R. and {Lazarz}, Daniel and {Leung}, Henry W. and {Leung}, Ho-Hin and {Lewis}, Hannah M. and {Li}, Cheng and {Li}, Ran and {Lian}, Jianhui and {Liang}, Fu-Heng and {Lin}, Lihwai and {Lin}, Yen-Ting and {Lin}, Sicheng and {Lintott}, Chris and {Long}, Dan and {Longa-Pe{\~n}a}, Pen{\'e}lope and {L{\'o}pez-Cob{\'a}}, Carlos and {Lu}, Shengdong and {Lundgren}, Britt F. and {Luo}, Yuanze and {Mackereth}, J. Ted and {de la Macorra}, Axel and {Mahadevan}, Suvrath and {Majewski}, Steven R. and {Manchado}, Arturo and {Mandeville}, Travis and {Maraston}, Claudia and {Margalef-Bentabol}, Berta and {Masseron}, Thomas and {Masters}, Karen L. and {Mathur}, Savita and {McDermid}, Richard M. and {Mckay}, Myles and {Merloni}, Andrea and {Merrifield}, Michael and {Meszaros}, Szabolcs and {Miglio}, Andrea and {Di Mille}, Francesco and {Minniti}, Dante and {Minsley}, Rebecca and {Monachesi}, Antonela},
        title = "{The Seventeenth Data Release of the Sloan Digital Sky Surveys: Complete Release of MaNGA, MaStar, and APOGEE-2 Data}",
      journal = {\apjs},
         year = 2022,
        month = apr,
       volume = {259},
       number = {2},
          eid = {35},
        pages = {35},
          doi = {10.3847/1538-4365/ac4414},
archivePrefix = {arXiv},
       eprint = {2112.02026},
 primaryClass = {astro-ph.GA},
       adsurl = {https://ui.adsabs.harvard.edu/abs/2022ApJS..259...35A}
}

@ARTICLE{Pasquali+10,
       author = {{Pasquali}, Anna and {Gallazzi}, Anna and {Fontanot}, Fabio and {van den Bosch}, Frank C. and {De Lucia}, Gabriella and {Mo}, H.~J. and {Yang}, Xiaohu},
        title = "{Ages and metallicities of central and satellite galaxies: implications for galaxy formation and evolution}",
      journal = {\mnras},
         year = 2010,
        month = sep,
       volume = {407},
       number = {2},
        pages = {937-954},
          doi = {10.1111/j.1365-2966.2010.17074.x},
archivePrefix = {arXiv},
       eprint = {0912.1853},
 primaryClass = {astro-ph.CO},
       adsurl = {https://ui.adsabs.harvard.edu/abs/2010MNRAS.407..937P}
}

@ARTICLE{Holmes+15,
       author = {{Holmes}, L. and {Spekkens}, K. and {S{\'a}nchez}, S.~F. and {Walcher}, C.~J. and {Garc{\'\i}a-Benito}, R. and {Mast}, D. and {Cortijo-Ferrero}, C. and {Kalinova}, V. and {Marino}, R.~A. and {Mendez-Abreu}, J. and {Barrera-Ballesteros}, J.~K.},
        title = "{The incidence of bar-like kinematic flows in CALIFA galaxies}",
      journal = {\mnras},
         year = 2015,
        month = aug,
       volume = {451},
       number = {4},
        pages = {4397-4411},
          doi = {10.1093/mnras/stv1254},
archivePrefix = {arXiv},
       eprint = {1506.01378},
 primaryClass = {astro-ph.GA},
       adsurl = {https://ui.adsabs.harvard.edu/abs/2015MNRAS.451.4397H}
}

@ARTICLE{Liu+13,
       author = {{Liu}, Guilin and {Zakamska}, Nadia L. and {Greene}, Jenny E. and {Nesvadba}, Nicole P.~H. and {Liu}, Xin},
        title = "{Observations of feedback from radio-quiet quasars - II. Kinematics of ionized gas nebulae}",
      journal = {\mnras},
         year = 2013,
        month = dec,
       volume = {436},
       number = {3},
        pages = {2576-2597},
          doi = {10.1093/mnras/stt1755},
archivePrefix = {arXiv},
       eprint = {1305.6922},
 primaryClass = {astro-ph.CO},
       adsurl = {https://ui.adsabs.harvard.edu/abs/2013MNRAS.436.2576L}
}

@ARTICLE{Kinemetry,
       author = {{Krajnovi{\'c}}, Davor and {Cappellari}, Michele and {de Zeeuw}, P. Tim and {Copin}, Yannick},
        title = "{Kinemetry: a generalization of photometry to the higher moments of the line-of-sight velocity distribution}",
      journal = {\mnras},
         year = 2006,
        month = mar,
       volume = {366},
       number = {3},
        pages = {787-802},
          doi = {10.1111/j.1365-2966.2005.09902.x},
archivePrefix = {arXiv},
       eprint = {astro-ph/0512200},
 primaryClass = {astro-ph},
       adsurl = {https://ui.adsabs.harvard.edu/abs/2006MNRAS.366..787K}
}

@ARTICLE{Feng+22,
       author = {{Feng}, Shuai and {Shen}, Shi-Yin and {Yuan}, Fang-Ting and {Dai}, Y. Sophia and {Masters}, Karen L.},
        title = "{The Velocity Map Asymmetry of Ionized Gas in MaNGA. I. The Catalog and General Properties}",
      journal = {\apjs},
         year = 2022,
        month = sep,
       volume = {262},
       number = {1},
          eid = {6},
        pages = {6},
          doi = {10.3847/1538-4365/ac80f2},
archivePrefix = {arXiv},
       eprint = {2207.06050},
 primaryClass = {astro-ph.GA},
       adsurl = {https://ui.adsabs.harvard.edu/abs/2022ApJS..262....6F}
}

@ARTICLE{Trussler+20,
       author = {{Trussler}, James and {Maiolino}, Roberto and {Maraston}, Claudia and {Peng}, Yingjie and {Thomas}, Daniel and {Goddard}, Daniel and {Lian}, Jianhui},
        title = "{Both starvation and outflows drive galaxy quenching}",
      journal = {\mnras},
         year = 2020,
        month = feb,
       volume = {491},
       number = {4},
        pages = {5406-5434},
          doi = {10.1093/mnras/stz3286},
archivePrefix = {arXiv},
       eprint = {1811.09283},
 primaryClass = {astro-ph.GA},
       adsurl = {https://ui.adsabs.harvard.edu/abs/2020MNRAS.491.5406T}
}

@ARTICLE{Rohr+24,
       author = {{Rohr}, Eric and {Pillepich}, Annalisa and {Nelson}, Dylan and {Ayromlou}, Mohammadreza and {Zinger}, Elad},
        title = "{The hot circumgalactic media of massive cluster satellites in the TNG-Cluster simulation: Existence and detectability}",
      journal = {\aap},
         year = 2024,
        month = jun,
       volume = {686},
          eid = {A86},
        pages = {A86},
          doi = {10.1051/0004-6361/202348583},
archivePrefix = {arXiv},
       eprint = {2311.06337},
 primaryClass = {astro-ph.GA},
       adsurl = {https://ui.adsabs.harvard.edu/abs/2024A&A...686A..86R}
}

@ARTICLE{Donnari+21b,
       author = {{Donnari}, Martina and {Pillepich}, Annalisa and {Joshi}, Gandhali D. and {Nelson}, Dylan and {Genel}, Shy and {Marinacci}, Federico and {Rodriguez-Gomez}, Vicente and {Pakmor}, R{\"u}diger and {Torrey}, Paul and {Vogelsberger}, Mark and {Hernquist}, Lars},
        title = "{Quenched fractions in the IllustrisTNG simulations: the roles of AGN feedback, environment, and pre-processing}",
      journal = {\mnras},
         year = 2021,
        month = jan,
       volume = {500},
       number = {3},
        pages = {4004-4024},
          doi = {10.1093/mnras/staa3006},
archivePrefix = {arXiv},
       eprint = {2008.00005},
 primaryClass = {astro-ph.GA},
       adsurl = {https://ui.adsabs.harvard.edu/abs/2021MNRAS.500.4004D}
}

@ARTICLE{Joshi+21,
       author = {{Joshi}, Gandhali D. and {Pillepich}, Annalisa and {Nelson}, Dylan and {Zinger}, Elad and {Marinacci}, Federico and {Springel}, Volker and {Vogelsberger}, Mark and {Hernquist}, Lars},
        title = "{The cumulative star formation histories of dwarf galaxies with TNG50. I: environment-driven diversity and connection to quenching}",
      journal = {\mnras},
         year = 2021,
        month = dec,
       volume = {508},
       number = {2},
        pages = {1652-1674},
          doi = {10.1093/mnras/stab2573},
archivePrefix = {arXiv},
       eprint = {2101.12226},
 primaryClass = {astro-ph.GA},
       adsurl = {https://ui.adsabs.harvard.edu/abs/2021MNRAS.508.1652J}
}

@ARTICLE{Lotz+19,
       author = {{Lotz}, Marcel and {Remus}, Rhea-Silvia and {Dolag}, Klaus and {Biviano}, Andrea and {Burkert}, Andreas},
        title = "{Gone after one orbit: How cluster environments quench galaxies}",
      journal = {\mnras},
         year = 2019,
        month = oct,
       volume = {488},
       number = {4},
        pages = {5370-5389},
          doi = {10.1093/mnras/stz2070},
archivePrefix = {arXiv},
       eprint = {1810.02382},
 primaryClass = {astro-ph.GA},
       adsurl = {https://ui.adsabs.harvard.edu/abs/2019MNRAS.488.5370L}
}

@ARTICLE{Bluck+20,
       author = {{Bluck}, Asa F.~L. and {Maiolino}, Roberto and {Piotrowska}, Joanna M. and {Trussler}, James and {Ellison}, Sara L. and {S{\'a}nchez}, Sebastian F. and {Thorp}, Mallory D. and {Teimoorinia}, Hossen and {Moreno}, Jorge and {Conselice}, Christopher J.},
        title = "{How do central and satellite galaxies quench? - Insights from spatially resolved spectroscopy in the MaNGA survey}",
      journal = {\mnras},
         year = 2020,
        month = nov,
       volume = {499},
       number = {1},
        pages = {230-268},
          doi = {10.1093/mnras/staa2806},
archivePrefix = {arXiv},
       eprint = {2009.05341},
 primaryClass = {astro-ph.GA},
       adsurl = {https://ui.adsabs.harvard.edu/abs/2020MNRAS.499..230B}
}

@ARTICLE{Weinberger+17a,
       author = {{Weinberger}, Rainer and {Ehlert}, Kristian and {Pfrommer}, Christoph and {Pakmor}, R{\"u}diger and {Springel}, Volker},
        title = "{Simulating the interaction of jets with the intracluster medium}",
      journal = {\mnras},
         year = 2017,
        month = oct,
       volume = {470},
       number = {4},
        pages = {4530-4546},
          doi = {10.1093/mnras/stx1409},
archivePrefix = {arXiv},
       eprint = {1703.09223},
 primaryClass = {astro-ph.GA},
       adsurl = {https://ui.adsabs.harvard.edu/abs/2017MNRAS.470.4530W}
}

@ARTICLE{Lim+25,
       author = {{Lim}, Seunghwan and {Tacchella}, Sandro and {Maiolino}, Roberto and {Schaye}, Joop and {Schaller}, Matthieu},
        title = "{In situ versus ex situ drivers of galaxy quenching: critical black hole mass and main sequence universality in the FLAMINGO simulation}",
      journal = {\mnras},
         year = 2025,
        month = nov,
       volume = {543},
       number = {3},
        pages = {2204-2221},
          doi = {10.1093/mnras/staf1578},
archivePrefix = {arXiv},
       eprint = {2504.02027},
 primaryClass = {astro-ph.GA},
       adsurl = {https://ui.adsabs.harvard.edu/abs/2025MNRAS.543.2204L}
}

@ARTICLE{Popesso+23,
       author = {{Popesso}, P. and {Concas}, A. and {Cresci}, G. and {Belli}, S. and {Rodighiero}, G. and {Inami}, H. and {Dickinson}, M. and {Ilbert}, O. and {Pannella}, M. and {Elbaz}, D.},
        title = "{The main sequence of star-forming galaxies across cosmic times}",
      journal = {\mnras},
         year = 2023,
        month = feb,
       volume = {519},
       number = {1},
        pages = {1526-1544},
          doi = {10.1093/mnras/stac3214},
archivePrefix = {arXiv},
       eprint = {2203.10487},
 primaryClass = {astro-ph.GA},
       adsurl = {https://ui.adsabs.harvard.edu/abs/2023MNRAS.519.1526P}
}

@ARTICLE{Wetzel+13,
       author = {{Wetzel}, Andrew R. and {Tinker}, Jeremy L. and {Conroy}, Charlie and {van den Bosch}, Frank C.},
        title = "{Galaxy evolution in groups and clusters: satellite star formation histories and quenching time-scales in a hierarchical Universe}",
      journal = {\mnras},
         year = 2013,
        month = jun,
       volume = {432},
       number = {1},
        pages = {336-358},
          doi = {10.1093/mnras/stt469},
archivePrefix = {arXiv},
       eprint = {1206.3571},
 primaryClass = {astro-ph.CO},
       adsurl = {https://ui.adsabs.harvard.edu/abs/2013MNRAS.432..336W}
}

@ARTICLE{Xie+20,
       author = {{Xie}, Lizhi and {De Lucia}, Gabriella and {Hirschmann}, Michaela and {Fontanot}, Fabio},
        title = "{The influence of environment on satellite galaxies in the GAEA semi-analytic model}",
      journal = {\mnras},
         year = 2020,
        month = nov,
       volume = {498},
       number = {3},
        pages = {4327-4344},
          doi = {10.1093/mnras/staa2370},
archivePrefix = {arXiv},
       eprint = {2003.12757},
 primaryClass = {astro-ph.GA},
       adsurl = {https://ui.adsabs.harvard.edu/abs/2020MNRAS.498.4327X}
}

@ARTICLE{Bower+06,
       author = {{Bower}, R.~G. and {Benson}, A.~J. and {Malbon}, R. and {Helly}, J.~C. and {Frenk}, C.~S. and {Baugh}, C.~M. and {Cole}, S. and {Lacey}, C.~G.},
        title = "{Breaking the hierarchy of galaxy formation}",
      journal = {\mnras},
         year = 2006,
        month = aug,
       volume = {370},
       number = {2},
        pages = {645-655},
          doi = {10.1111/j.1365-2966.2006.10519.x},
archivePrefix = {arXiv},
       eprint = {astro-ph/0511338},
 primaryClass = {astro-ph},
       adsurl = {https://ui.adsabs.harvard.edu/abs/2006MNRAS.370..645B}
}

@ARTICLE{Visser-Zadvornyi+25,
       author = {{Visser-Zadvornyi}, Anatolii I. and {Carstairs}, Mary E. and {Oman}, Kyle A. and {Verheijen}, Marc A.~W.},
        title = "{Star formation and stellar \& AGN feedback in the absence of accretion, not gas stripping, set the quenching time-scale in satellite galaxies}",
      journal = {\mnras},
         year = 2025,
        month = jun,
       volume = {540},
       number = {2},
        pages = {1730-1744},
          doi = {10.1093/mnras/staf802},
archivePrefix = {arXiv},
       eprint = {2503.15183},
 primaryClass = {astro-ph.GA},
       adsurl = {https://ui.adsabs.harvard.edu/abs/2025MNRAS.540.1730V}
}

@ARTICLE{Donnari+21,
       author = {{Donnari}, Martina and {Pillepich}, Annalisa and {Joshi}, Gandhali D. and {Nelson}, Dylan and {Genel}, Shy and {Marinacci}, Federico and {Rodriguez-Gomez}, Vicente and {Pakmor}, R{\"u}diger and {Torrey}, Paul and {Vogelsberger}, Mark and {Hernquist}, Lars},
        title = "{Quenched fractions in the IllustrisTNG simulations: the roles of AGN feedback, environment, and pre-processing}",
      journal = {\mnras},
         year = 2021,
        month = jan,
       volume = {500},
       number = {3},
        pages = {4004-4024},
          doi = {10.1093/mnras/staa3006},
archivePrefix = {arXiv},
       eprint = {2008.00005},
 primaryClass = {astro-ph.GA},
       adsurl = {https://ui.adsabs.harvard.edu/abs/2021MNRAS.500.4004D}
}

@ARTICLE{Peng+10,
       author = {{Peng}, Ying-jie and {Lilly}, Simon J. and {Kova{\v{c}}}, Katarina and {Bolzonella}, Micol and {Pozzetti}, Lucia and {Renzini}, Alvio and {Zamorani}, Gianni and {Ilbert}, Olivier and {Knobel}, Christian and {Iovino}, Angela and {Maier}, Christian and {Cucciati}, Olga and {Tasca}, Lidia and {Carollo}, C. Marcella and {Silverman}, John and {Kampczyk}, Pawel and {de Ravel}, Loic and {Sanders}, David and {Scoville}, Nicholas and {Contini}, Thierry and {Mainieri}, Vincenzo and {Scodeggio}, Marco and {Kneib}, Jean-Paul and {Le F{\`e}vre}, Olivier and {Bardelli}, Sandro and {Bongiorno}, Angela and {Caputi}, Karina and {Coppa}, Graziano and {de la Torre}, Sylvain and {Franzetti}, Paolo and {Garilli}, Bianca and {Lamareille}, Fabrice and {Le Borgne}, Jean-Francois and {Le Brun}, Vincent and {Mignoli}, Marco and {Perez Montero}, Enrique and {Pello}, Roser and {Ricciardelli}, Elena and {Tanaka}, Masayuki and {Tresse}, Laurence and {Vergani}, Daniela and {Welikala}, Niraj and {Zucca}, Elena and {Oesch}, Pascal and {Abbas}, Ummi and {Barnes}, Luke and {Bordoloi}, Rongmon and {Bottini}, Dario and {Cappi}, Alberto and {Cassata}, Paolo and {Cimatti}, Andrea and {Fumana}, Marco and {Hasinger}, Gunther and {Koekemoer}, Anton and {Leauthaud}, Alexei and {Maccagni}, Dario and {Marinoni}, Christian and {McCracken}, Henry and {Memeo}, Pierdomenico and {Meneux}, Baptiste and {Nair}, Preethi and {Porciani}, Cristiano and {Presotto}, Valentina and {Scaramella}, Roberto},
        title = "{Mass and Environment as Drivers of Galaxy Evolution in SDSS and zCOSMOS and the Origin of the Schechter Function}",
      journal = {\apj},
         year = 2010,
        month = sep,
       volume = {721},
       number = {1},
        pages = {193-221},
          doi = {10.1088/0004-637X/721/1/193},
archivePrefix = {arXiv},
       eprint = {1003.4747},
 primaryClass = {astro-ph.CO},
       adsurl = {https://ui.adsabs.harvard.edu/abs/2010ApJ...721..193P}
}

@ARTICLE{Bell+04,
       author = {{Bell}, Eric F. and {Wolf}, Christian and {Meisenheimer}, Klaus and {Rix}, Hans-Walter and {Borch}, Andrea and {Dye}, Simon and {Kleinheinrich}, Martina and {Wisotzki}, Lutz and {McIntosh}, Daniel H.},
        title = "{Nearly 5000 Distant Early-Type Galaxies in COMBO-17: A Red Sequence and Its Evolution since z\raisebox{-0.5ex}\textasciitilde1}",
      journal = {\apj},
         year = 2004,
        month = jun,
       volume = {608},
       number = {2},
        pages = {752-767},
          doi = {10.1086/420778},
archivePrefix = {arXiv},
       eprint = {astro-ph/0303394},
 primaryClass = {astro-ph},
       adsurl = {https://ui.adsabs.harvard.edu/abs/2004ApJ...608..752B}
}

@ARTICLE{Peng+15,
       author = {{Peng}, Y. and {Maiolino}, R. and {Cochrane}, R.},
        title = "{Strangulation as the primary mechanism for shutting down star formation in galaxies}",
      journal = {\nat},
         year = 2015,
        month = may,
       volume = {521},
       number = {7551},
        pages = {192-195},
          doi = {10.1038/nature14439},
archivePrefix = {arXiv},
       eprint = {1505.03143},
 primaryClass = {astro-ph.GA},
       adsurl = {https://ui.adsabs.harvard.edu/abs/2015Natur.521..192P}
}

@ARTICLE{pyPipe3D,
       author = {{S{\'a}nchez}, S.~F. and {Barrera-Ballesteros}, J.~K. and {Lacerda}, E. and {Mej{\'\i}a-Narvaez}, A. and {Camps-Fari{\~n}a}, A. and {Bruzual}, Gustavo and {Espinosa-Ponce}, C. and {Rodr{\'\i}guez-Puebla}, A. and {Calette}, A.~R. and {Ibarra-Medel}, H. and {Avila-Reese}, V. and {Hernandez-Toledo}, H. and {Bershady}, M.~A. and {Cano-Diaz}, M. and {Munguia-Cordova}, A.~M.},
        title = "{SDSS-IV MaNGA: pyPipe3D Analysis Release for 10,000 Galaxies}",
      journal = {\apjs},
         year = 2022,
        month = oct,
       volume = {262},
       number = {2},
          eid = {36},
        pages = {36},
          doi = {10.3847/1538-4365/ac7b8f},
archivePrefix = {arXiv},
       eprint = {2206.07062},
 primaryClass = {astro-ph.GA},
       adsurl = {https://ui.adsabs.harvard.edu/abs/2022ApJS..262...36S}
}

@ARTICLE{Cortese+21,
       author = {{Cortese}, L. and {Catinella}, B. and {Smith}, R.},
        title = "{The Dawes Review 9: The role of cold gas stripping on the star formation quenching of satellite galaxies}",
      journal = {\pasa},
         year = 2021,
        month = aug,
       volume = {38},
          eid = {e035},
        pages = {e035},
          doi = {10.1017/pasa.2021.18},
archivePrefix = {arXiv},
       eprint = {2104.02193},
 primaryClass = {astro-ph.GA},
       adsurl = {https://ui.adsabs.harvard.edu/abs/2021PASA...38...35C}
}

@ARTICLE{Cortese+19,
       author = {{Cortese}, L. and {van de Sande}, J. and {Lagos}, C.~P. and {Catinella}, B. and {Davies}, L.~J.~M. and {Croom}, S.~M. and {Brough}, S. and {Bryant}, J.~J. and {Lawrence}, J.~S. and {Owers}, M.~S. and {Richards}, S.~N. and {Sweet}, S.~M. and {Bland-Hawthorn}, J.},
        title = "{The SAMI Galaxy Survey: satellite galaxies undergo little structural change during their quenching phase}",
      journal = {\mnras},
         year = 2019,
        month = may,
       volume = {485},
       number = {2},
        pages = {2656-2665},
          doi = {10.1093/mnras/stz485},
archivePrefix = {arXiv},
       eprint = {1902.05652},
 primaryClass = {astro-ph.GA},
       adsurl = {https://ui.adsabs.harvard.edu/abs/2019MNRAS.485.2656C}
}

@ARTICLE{Gunn+72,
       author = {{Gunn}, James E. and {Gott}, III, J. Richard},
        title = "{On the Infall of Matter Into Clusters of Galaxies and Some Effects on Their Evolution}",
      journal = {\apj},
         year = 1972,
        month = aug,
       volume = {176},
        pages = {1},
          doi = {10.1086/151605},
       adsurl = {https://ui.adsabs.harvard.edu/abs/1972ApJ...176....1G}
}

@ARTICLE{Wetzel+15,
       author = {{Wetzel}, Andrew R. and {Tollerud}, Erik J. and {Weisz}, Daniel R.},
        title = "{Rapid Environmental Quenching of Satellite Dwarf Galaxies in the Local Group}",
      journal = {\apjl},
         year = 2015,
        month = jul,
       volume = {808},
       number = {1},
          eid = {L27},
        pages = {L27},
          doi = {10.1088/2041-8205/808/1/L27},
archivePrefix = {arXiv},
       eprint = {1503.06799},
 primaryClass = {astro-ph.GA},
       adsurl = {https://ui.adsabs.harvard.edu/abs/2015ApJ...808L..27W}
}

@ARTICLE{Jaffe+15,
       author = {{Jaff{\'e}}, Yara L. and {Smith}, Rory and {Candlish}, Graeme N. and {Poggianti}, Bianca M. and {Sheen}, Yun-Kyeong and {Verheijen}, Marc A.~W.},
        title = "{BUDHIES II: a phase-space view of H I gas stripping and star formation quenching in cluster galaxies}",
      journal = {\mnras},
         year = 2015,
        month = apr,
       volume = {448},
       number = {2},
        pages = {1715-1728},
          doi = {10.1093/mnras/stv100},
archivePrefix = {arXiv},
       eprint = {1501.03819},
 primaryClass = {astro-ph.GA},
       adsurl = {https://ui.adsabs.harvard.edu/abs/2015MNRAS.448.1715J}
}

@ARTICLE{Muzzin+14,
       author = {{Muzzin}, Adam and {van der Burg}, R.~F.~J. and {McGee}, Sean L. and {Balogh}, Michael and {Franx}, Marijn and {Hoekstra}, Henk and {Hudson}, Michael J. and {Noble}, Allison and {Taranu}, Dan S. and {Webb}, Tracy and {Wilson}, Gillian and {Yee}, H.~K.~C.},
        title = "{The Phase Space and Stellar Populations of Cluster Galaxies at z \raisebox{-0.5ex}\textasciitilde 1: Simultaneous Constraints on the Location and Timescale of Satellite Quenching}",
      journal = {\apj},
         year = 2014,
        month = nov,
       volume = {796},
       number = {1},
          eid = {65},
        pages = {65},
          doi = {10.1088/0004-637X/796/1/65},
archivePrefix = {arXiv},
       eprint = {1402.7077},
 primaryClass = {astro-ph.GA},
       adsurl = {https://ui.adsabs.harvard.edu/abs/2014ApJ...796...65M}
}

@ARTICLE{Popesso+15,
       author = {{Popesso}, P. and {Biviano}, A. and {Finoguenov}, A. and {Wilman}, D. and {Salvato}, M. and {Magnelli}, B. and {Gruppioni}, C. and {Pozzi}, F. and {Rodighiero}, G. and {Ziparo}, F. and {Berta}, S. and {Elbaz}, D. and {Dickinson}, M. and {Lutz}, D. and {Altieri}, B. and {Aussel}, H. and {Cimatti}, A. and {Fadda}, D. and {Ilbert}, O. and {Le Floch}, E. and {Nordon}, R. and {Poglitsch}, A. and {Xu}, C.~K.},
        title = "{The evolution of galaxy star formation activity in massive haloes}",
      journal = {\aap},
         year = 2015,
        month = feb,
       volume = {574},
          eid = {A105},
        pages = {A105},
          doi = {10.1051/0004-6361/201424711},
archivePrefix = {arXiv},
       eprint = {1407.8214},
 primaryClass = {astro-ph.GA},
       adsurl = {https://ui.adsabs.harvard.edu/abs/2015A&A...574A.105P}
}

@ARTICLE{Behroozi+19,
       author = {{Behroozi}, Peter and {Wechsler}, Risa H. and {Hearin}, Andrew P. and {Conroy}, Charlie},
        title = "{UNIVERSEMACHINE: The correlation between galaxy growth and dark matter halo assembly from z = 0-10}",
      journal = {\mnras},
         year = 2019,
        month = sep,
       volume = {488},
       number = {3},
        pages = {3143-3194},
          doi = {10.1093/mnras/stz1182},
archivePrefix = {arXiv},
       eprint = {1806.07893},
 primaryClass = {astro-ph.GA},
       adsurl = {https://ui.adsabs.harvard.edu/abs/2019MNRAS.488.3143B}
}

@ARTICLE{Croton+06,
       author = {{Croton}, Darren J. and {Springel}, Volker and {White}, Simon D.~M. and {De Lucia}, G. and {Frenk}, C.~S. and {Gao}, L. and {Jenkins}, A. and {Kauffmann}, G. and {Navarro}, J.~F. and {Yoshida}, N.},
        title = "{The many lives of active galactic nuclei: cooling flows, black holes and the luminosities and colours of galaxies}",
      journal = {\mnras},
         year = 2006,
        month = jan,
       volume = {365},
       number = {1},
        pages = {11-28},
          doi = {10.1111/j.1365-2966.2005.09675.x},
archivePrefix = {arXiv},
       eprint = {astro-ph/0508046},
 primaryClass = {astro-ph},
       adsurl = {https://ui.adsabs.harvard.edu/abs/2006MNRAS.365...11C}
}

@ARTICLE{Oman+16,
       author = {{Oman}, Kyle A. and {Hudson}, Michael J.},
        title = "{Satellite quenching time-scales in clusters from projected phase space measurements matched to simulated orbits}",
      journal = {\mnras},
         year = 2016,
        month = dec,
       volume = {463},
       number = {3},
        pages = {3083-3095},
          doi = {10.1093/mnras/stw2195},
archivePrefix = {arXiv},
       eprint = {1607.07934},
 primaryClass = {astro-ph.GA},
       adsurl = {https://ui.adsabs.harvard.edu/abs/2016MNRAS.463.3083O}
}

@ARTICLE{Baxter+25,
       author = {{Baxter}, Devontae C. and {Fillingham}, Sean P. and {Coil}, Alison L. and {Cooper}, Michael C.},
        title = "{The Importance of Gas Starvation in Driving Satellite Quenching in Galaxy Groups at z \raisebox{-0.5ex}\textasciitilde 0.8}",
      journal = {\apj},
         year = 2025,
        month = jan,
       volume = {979},
       number = {1},
          eid = {41},
        pages = {41},
          doi = {10.3847/1538-4357/ad9aa4},
archivePrefix = {arXiv},
       eprint = {2412.02766},
 primaryClass = {astro-ph.GA},
       adsurl = {https://ui.adsabs.harvard.edu/abs/2025ApJ...979...41B}
}

@ARTICLE{Pasquali+19,
       author = {{Pasquali}, A. and {Smith}, R. and {Gallazzi}, A. and {De Lucia}, G. and {Zibetti}, S. and {Hirschmann}, M. and {Yi}, S.~K.},
        title = "{Physical properties of SDSS satellite galaxies in projected phase space}",
      journal = {\mnras},
         year = 2019,
        month = apr,
       volume = {484},
       number = {2},
        pages = {1702-1723},
          doi = {10.1093/mnras/sty3530},
archivePrefix = {arXiv},
       eprint = {1901.04238},
 primaryClass = {astro-ph.GA},
       adsurl = {https://ui.adsabs.harvard.edu/abs/2019MNRAS.484.1702P}
}

@ARTICLE{Genel+16,
       author = {{Genel}, Shy},
        title = "{How Environment Affects Galaxy Metallicity through Stripping and Formation History: Lessons from the Illustris Simulation}",
      journal = {\apj},
         year = 2016,
        month = may,
       volume = {822},
       number = {2},
          eid = {107},
        pages = {107},
          doi = {10.3847/0004-637X/822/2/107},
archivePrefix = {arXiv},
       eprint = {1602.02773},
 primaryClass = {astro-ph.GA},
       adsurl = {https://ui.adsabs.harvard.edu/abs/2016ApJ...822..107G}
}

@ARTICLE{Peroux+20,
       author = {{P{\'e}roux}, C{\'e}line and {Howk}, J. Christopher},
        title = "{The Cosmic Baryon and Metal Cycles}",
      journal = {\araa},
         year = 2020,
        month = aug,
       volume = {58},
        pages = {363-406},
          doi = {10.1146/annurev-astro-021820-120014},
archivePrefix = {arXiv},
       eprint = {2011.01935},
 primaryClass = {astro-ph.GA},
       adsurl = {https://ui.adsabs.harvard.edu/abs/2020ARA&A..58..363P}
}

@ARTICLE{McCarthy+08,
       author = {{McCarthy}, I.~G. and {Frenk}, C.~S. and {Font}, A.~S. and {Lacey}, C.~G. and {Bower}, R.~G. and {Mitchell}, N.~L. and {Balogh}, M.~L. and {Theuns}, T.},
        title = "{Ram pressure stripping the hot gaseous haloes of galaxies in groups and clusters}",
      journal = {\mnras},
         year = 2008,
        month = jan,
       volume = {383},
       number = {2},
        pages = {593-605},
          doi = {10.1111/j.1365-2966.2007.12577.x},
archivePrefix = {arXiv},
       eprint = {0710.0964},
 primaryClass = {astro-ph},
       adsurl = {https://ui.adsabs.harvard.edu/abs/2008MNRAS.383..593M}
}

@ARTICLE{Bahe+15,
       author = {{Bah{\'e}}, Yannick M. and {McCarthy}, Ian G.},
        title = "{Star formation quenching in simulated group and cluster galaxies: when, how, and why?}",
      journal = {\mnras},
         year = 2015,
        month = feb,
       volume = {447},
       number = {1},
        pages = {969-992},
          doi = {10.1093/mnras/stu2293},
archivePrefix = {arXiv},
       eprint = {1410.8161},
 primaryClass = {astro-ph.GA},
       adsurl = {https://ui.adsabs.harvard.edu/abs/2015MNRAS.447..969B}
}

@ARTICLE{Davies+20,
       author = {{Davies}, Jonathan J. and {Crain}, Robert A. and {Oppenheimer}, Benjamin D. and {Schaye}, Joop},
        title = "{The quenching and morphological evolution of central galaxies is facilitated by the feedback-driven expulsion of circumgalactic gas}",
      journal = {\mnras},
         year = 2020,
        month = jan,
       volume = {491},
       number = {3},
        pages = {4462-4480},
          doi = {10.1093/mnras/stz3201},
archivePrefix = {arXiv},
       eprint = {1908.11380},
 primaryClass = {astro-ph.GA},
       adsurl = {https://ui.adsabs.harvard.edu/abs/2020MNRAS.491.4462D}
}

@ARTICLE{Schaefer+17,
       author = {{Schaefer}, A.~L. and {Croom}, S.~M. and {Allen}, J.~T. and {Brough}, S. and {Medling}, A.~M. and {Ho}, I.-T. and {Scott}, N. and {Richards}, S.~N. and {Pracy}, M.~B. and {Gunawardhana}, M.~L.~P. and {Norberg}, P. and {Alpaslan}, M. and {Bauer}, A.~E. and {Bekki}, K. and {Bland-Hawthorn}, J. and {Bloom}, J.~V. and {Bryant}, J.~J. and {Couch}, W.~J. and {Driver}, S.~P. and {Fogarty}, L.~M.~R. and {Foster}, C. and {Goldstein}, G. and {Green}, A.~W. and {Hopkins}, A.~M. and {Konstantopoulos}, I.~S. and {Lawrence}, J.~S. and {L{\'o}pez-S{\'a}nchez}, A.~R. and {Lorente}, N.~P.~F. and {Owers}, M.~S. and {Sharp}, R. and {Sweet}, S.~M. and {Taylor}, E.~N. and {van de Sande}, J. and {Walcher}, C.~J. and {Wong}, O.~I.},
        title = "{The SAMI Galaxy Survey: spatially resolving the environmental quenching of star formation in GAMA galaxies}",
      journal = {\mnras},
         year = 2017,
        month = jan,
       volume = {464},
       number = {1},
        pages = {121-142},
          doi = {10.1093/mnras/stw2289},
archivePrefix = {arXiv},
       eprint = {1609.02635},
 primaryClass = {astro-ph.GA},
       adsurl = {https://ui.adsabs.harvard.edu/abs/2017MNRAS.464..121S}
}

@ARTICLE{Tumlinson+17,
       author = {{Tumlinson}, Jason and {Peeples}, Molly S. and {Werk}, Jessica K.},
        title = "{The Circumgalactic Medium}",
      journal = {\araa},
         year = 2017,
        month = aug,
       volume = {55},
       number = {1},
        pages = {389-432},
          doi = {10.1146/annurev-astro-091916-055240},
archivePrefix = {arXiv},
       eprint = {1709.09180},
 primaryClass = {astro-ph.GA},
       adsurl = {https://ui.adsabs.harvard.edu/abs/2017ARA&A..55..389T}
}

@ARTICLE{Oman+21,
       author = {{Oman}, Kyle A. and {Bah{\'e}}, Yannick M. and {Healy}, Julia and {Hess}, Kelley M. and {Hudson}, Michael J. and {Verheijen}, Marc A.~W.},
        title = "{A homogeneous measurement of the delay between the onsets of gas stripping and star formation quenching in satellite galaxies of groups and clusters}",
      journal = {\mnras},
         year = 2021,
        month = mar,
       volume = {501},
       number = {4},
        pages = {5073-5095},
          doi = {10.1093/mnras/staa3845},
archivePrefix = {arXiv},
       eprint = {2009.00667},
 primaryClass = {astro-ph.GA},
       adsurl = {https://ui.adsabs.harvard.edu/abs/2021MNRAS.501.5073O}
}

@ARTICLE{Popesso+19a,
       author = {{Popesso}, P. and {Concas}, A. and {Morselli}, L. and {Schreiber}, C. and {Rodighiero}, G. and {Cresci}, G. and {Belli}, S. and {Erfanianfar}, G. and {Mancini}, C. and {Inami}, H. and {Dickinson}, M. and {Ilbert}, O. and {Pannella}, M. and {Elbaz}, D.},
        title = "{The main sequence of star-forming galaxies - I. The local relation and its bending}",
      journal = {\mnras},
         year = 2019,
        month = mar,
       volume = {483},
       number = {3},
        pages = {3213-3226},
          doi = {10.1093/mnras/sty3210},
archivePrefix = {arXiv},
       eprint = {1812.07057},
 primaryClass = {astro-ph.GA},
       adsurl = {https://ui.adsabs.harvard.edu/abs/2019MNRAS.483.3213P}
}

@ARTICLE{Gallazzi+21,
       author = {{Gallazzi}, Anna R. and {Pasquali}, A. and {Zibetti}, S. and {Barbera}, F. La},
        title = "{Galaxy evolution across environments as probed by the ages, stellar metallicities, and [{\ensuremath{\alpha}} /Fe] of central and satellite galaxies}",
      journal = {\mnras},
         year = 2021,
        month = apr,
       volume = {502},
       number = {3},
        pages = {4457-4478},
          doi = {10.1093/mnras/stab265},
archivePrefix = {arXiv},
       eprint = {2010.04733},
 primaryClass = {astro-ph.GA},
       adsurl = {https://ui.adsabs.harvard.edu/abs/2021MNRAS.502.4457G}
}

@ARTICLE{Westfall+19,
       author = {{Westfall}, Kyle B. and {Cappellari}, Michele and {Bershady}, Matthew A. and {Bundy}, Kevin and {Belfiore}, Francesco and {Ji}, Xihan and {Law}, David R. and {Schaefer}, Adam and {Shetty}, Shravan and {Tremonti}, Christy A. and {Yan}, Renbin and {Andrews}, Brett H. and {Brownstein}, Joel R. and {Cherinka}, Brian and {Coccato}, Lodovico and {Drory}, Niv and {Maraston}, Claudia and {Parikh}, Taniya and {S{\'a}nchez-Gallego}, Jos{\'e} R. and {Thomas}, Daniel and {Weijmans}, Anne-Marie and {Barrera-Ballesteros}, Jorge and {Du}, Cheng and {Goddard}, Daniel and {Li}, Niu and {Masters}, Karen and {Ibarra Medel}, H{\'e}ctor Javier and {S{\'a}nchez}, Sebasti{\'a}n F. and {Yang}, Meng and {Zheng}, Zheng and {Zhou}, Shuang},
        title = "{The Data Analysis Pipeline for the SDSS-IV MaNGA IFU Galaxy Survey: Overview}",
      journal = {\aj},
         year = 2019,
        month = dec,
       volume = {158},
       number = {6},
          eid = {231},
        pages = {231},
          doi = {10.3847/1538-3881/ab44a2},
archivePrefix = {arXiv},
       eprint = {1901.00856},
 primaryClass = {astro-ph.GA},
       adsurl = {https://ui.adsabs.harvard.edu/abs/2019AJ....158..231W}
}

@ARTICLE{Larson+80,
       author = {{Larson}, R.~B. and {Tinsley}, B.~M. and {Caldwell}, C.~N.},
        title = "{The evolution of disk galaxies and the origin of S0 galaxies}",
      journal = {\apj},
         year = 1980,
        month = may,
       volume = {237},
        pages = {692-707},
          doi = {10.1086/157917},
       adsurl = {https://ui.adsabs.harvard.edu/abs/1980ApJ...237..692L}
}

@ARTICLE{Tacchella+16,
       author = {{Tacchella}, Sandro and {Dekel}, Avishai and {Carollo}, C. Marcella and {Ceverino}, Daniel and {DeGraf}, Colin and {Lapiner}, Sharon and {Mandelker}, Nir and {Primack}, Joel R.},
        title = "{Evolution of density profiles in high-z galaxies: compaction and quenching inside-out}",
      journal = {\mnras},
         year = 2016,
        month = may,
       volume = {458},
       number = {1},
        pages = {242-263},
          doi = {10.1093/mnras/stw303},
archivePrefix = {arXiv},
       eprint = {1509.00017},
 primaryClass = {astro-ph.GA},
       adsurl = {https://ui.adsabs.harvard.edu/abs/2016MNRAS.458..242T}
}

@ARTICLE{Tacchella+15,
       author = {{Tacchella}, S. and {Carollo}, C.~M. and {Renzini}, A. and {F{\"o}rster Schreiber}, N.~M. and {Lang}, P. and {Wuyts}, S. and {Cresci}, G. and {Dekel}, A. and {Genzel}, R. and {Lilly}, S.~J. and {Mancini}, C. and {Newman}, S. and {Onodera}, M. and {Shapley}, A. and {Tacconi}, L. and {Woo}, J. and {Zamorani}, G.},
        title = "{Evidence for mature bulges and an inside-out quenching phase 3 billion years after the Big Bang}",
      journal = {Science},
         year = 2015,
        month = apr,
       volume = {348},
       number = {6232},
        pages = {314-317},
          doi = {10.1126/science.1261094},
archivePrefix = {arXiv},
       eprint = {1504.04021},
 primaryClass = {astro-ph.GA},
       adsurl = {https://ui.adsabs.harvard.edu/abs/2015Sci...348..314T}
}

\begin{appendix}
\onecolumn

\section{Sample distribution in $M_\star$, $M_{\rm halo}$, and SFR}\label{AppendixA}
\begin{figure*}[h!]
    \centering
    \includegraphics[width=\textwidth]{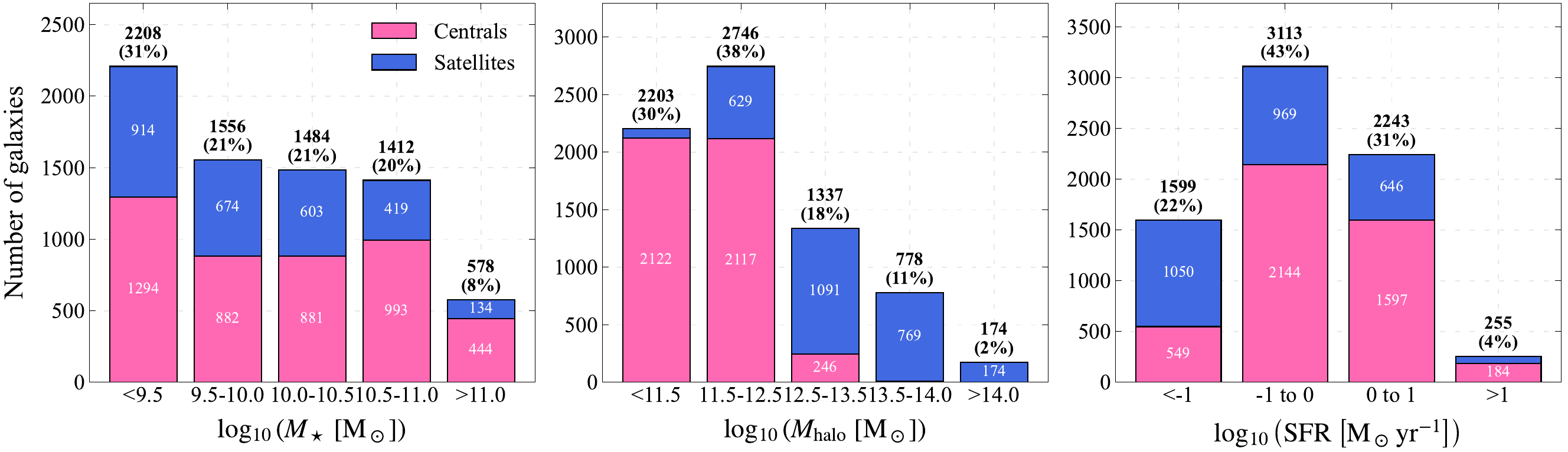}
    \caption{Distribution of galaxies in our sample as a function of stellar mass (\textit{left~panel}), host halo mass (\textit{middle~panel}), and star formation rate (\textit{right~panel}). Satellites are shown in blue and centrals in pink.}
    \label{fig:AppendixA}
\end{figure*}

\section{Dependence of CGM-depletion timescale on stellar mass}\label{AppendixB}

\begin{figure*}[h!]
    \centering
    \includegraphics[width=0.9\textwidth]{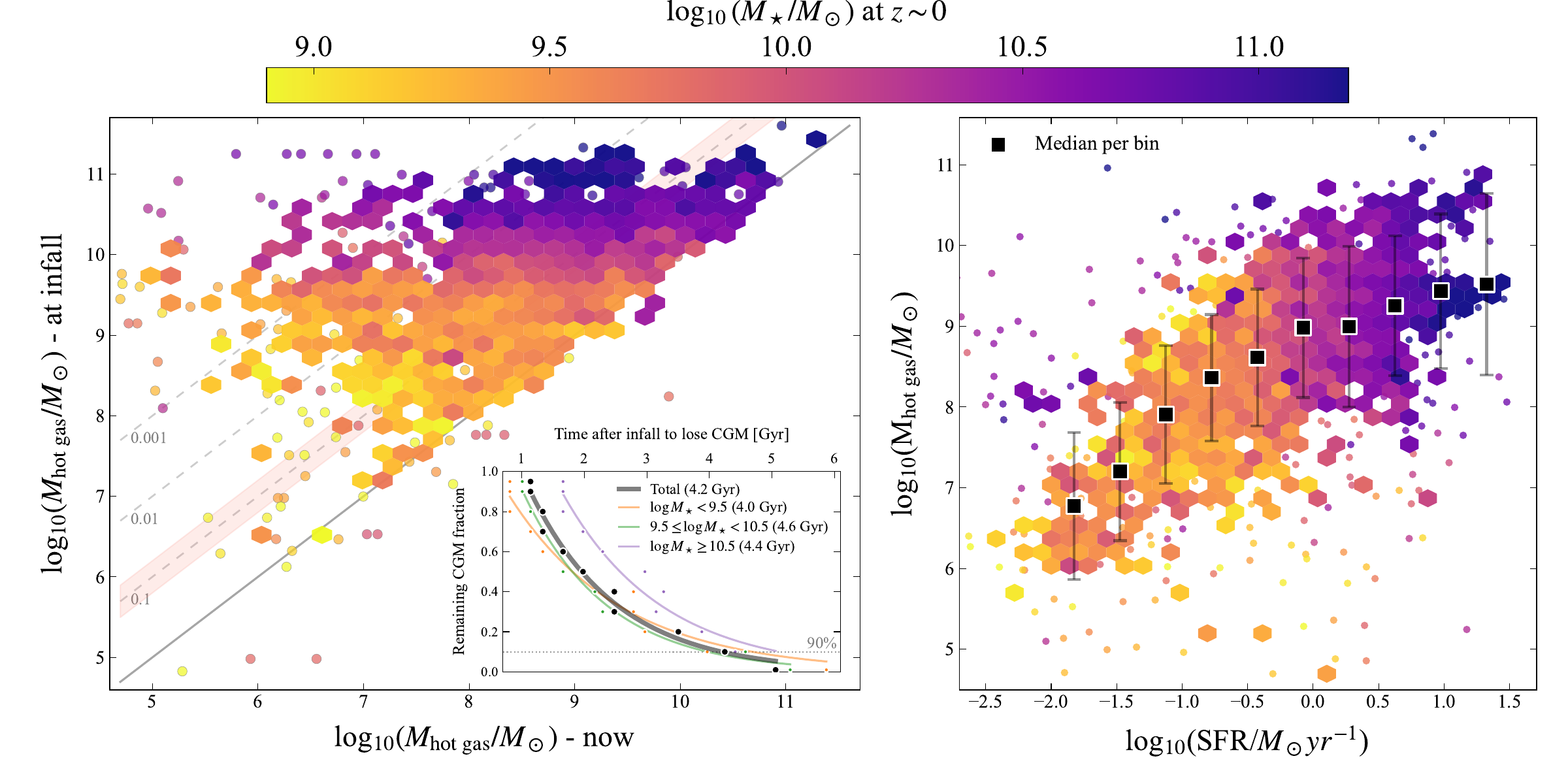}
    \caption{Same as Fig.\,\ref{fig:MainPlot}, but color-coded by the present-day ($z$$\,\sim\,$$0$) stellar mass. \textit{Left:} hot gas mass at $z$$\,\sim\,$$0$ versus at infall. The inset shows the remaining hot CGM fraction as a function of time after infall for the full satellite sample (black) and separately for the three stellar-mass bins, with the $90\%$ gas-loss timescale, $t_{90}$, indicated for each. The $t_{90}$ values agree across stellar-mass bins, showing that the CGM-depletion timescale is independent of stellar mass. \textit{Right:} hot gas mass versus SFR. Unlike the \textit{left panel}, $M_{\rm hot\,gas}$--$\rm SFR$ shows a dependence on stellar mass.}
    \label{fig:Mstar_dependence}
\end{figure*}

To check that the trends presented in $\S$\ref{results}, particularly in $\S$\ref{hot_gas}, reflect environmental (infall-driven) processes rather than a dependence on galaxy stellar mass, we repeat the analysis with the satellite sample color-coded by the present-day ($z$$\,\sim\,$$0$) stellar mass. Furthermore, we split the sample into three $\log_{10}(M_\star/\rm M_\odot)$ bins: ${<}9.5$, $9.5$–$10.5$, and ${\geq}10.5$. The result is shown in Fig.\,\ref{fig:Mstar_dependence}.

As expected from the stellar-to-halo-mass relation, stellar mass correlates smoothly with the absolute hot-gas content: more massive satellites bring in, and retain, larger CGM reservoirs, producing the monotonic color gradient along the $M_{\rm hot,\,now}$--$M_{\rm hot,\,infall}$ sequence (Fig.\,\ref{fig:Mstar_dependence}, left). Crucially, this gradient runs along the sequence rather than across it: at fixed stellar mass, satellites span the full range of CGM depletion, from systems near the 1:1 line to those that have lost more than $99\%$ of their hot gas. Stellar mass therefore sets the normalization of the reservoir, not the degree to which it is stripped.

We quantify this by measuring the time required to remove $90\%$ of the CGM, $t_{90}$, in each stellar-mass bin (inset of Fig.\,\ref{fig:Mstar_dependence}, left). We obtain $t_{90} = 4.0^{+1.0}_{-1.0}$, $4.6^{+0.8}_{-0.8}$, and $4.4^{+0.8}_{-0.8}\,\mathrm{Gyr}$ for the low-, intermediate-, and high-mass bins, respectively. We stress that all these three estimates are consistent within their $3\sigma$ uncertainties with the value derived for the full sample ($t_{90}\simeq4.2^{+0.6}_{-0.6}\,\mathrm{Gyr}$; $\S$\ref{hot_gas}). Therefore, we conclude that the CGM-depletion timescale we present in this work shows no significant dependence on stellar mass.

For completeness, the right-hand panel of Fig.\,\ref{fig:Mstar_dependence} shows that the $M_{\rm hot}$--SFR relation carries a residual stellar-mass trend, as expected since both quantities scale jointly with mass.

\twocolumn
\section{Cool gas evolution and depletion timescales}\label{AppendixC}
For every satellite we trace the bound cool-gas mass ($T<3\times10^{4}$\,K) along its main-progenitor branch in the \textsc{SubLink} tree and
interpolate it onto a uniform grid in time since first infall, using the same approach
applied to the hot component described in $\S$\ref{hotgasmass}.

\begin{figure}[h!]
    \centering
    \includegraphics[width=0.48\textwidth]{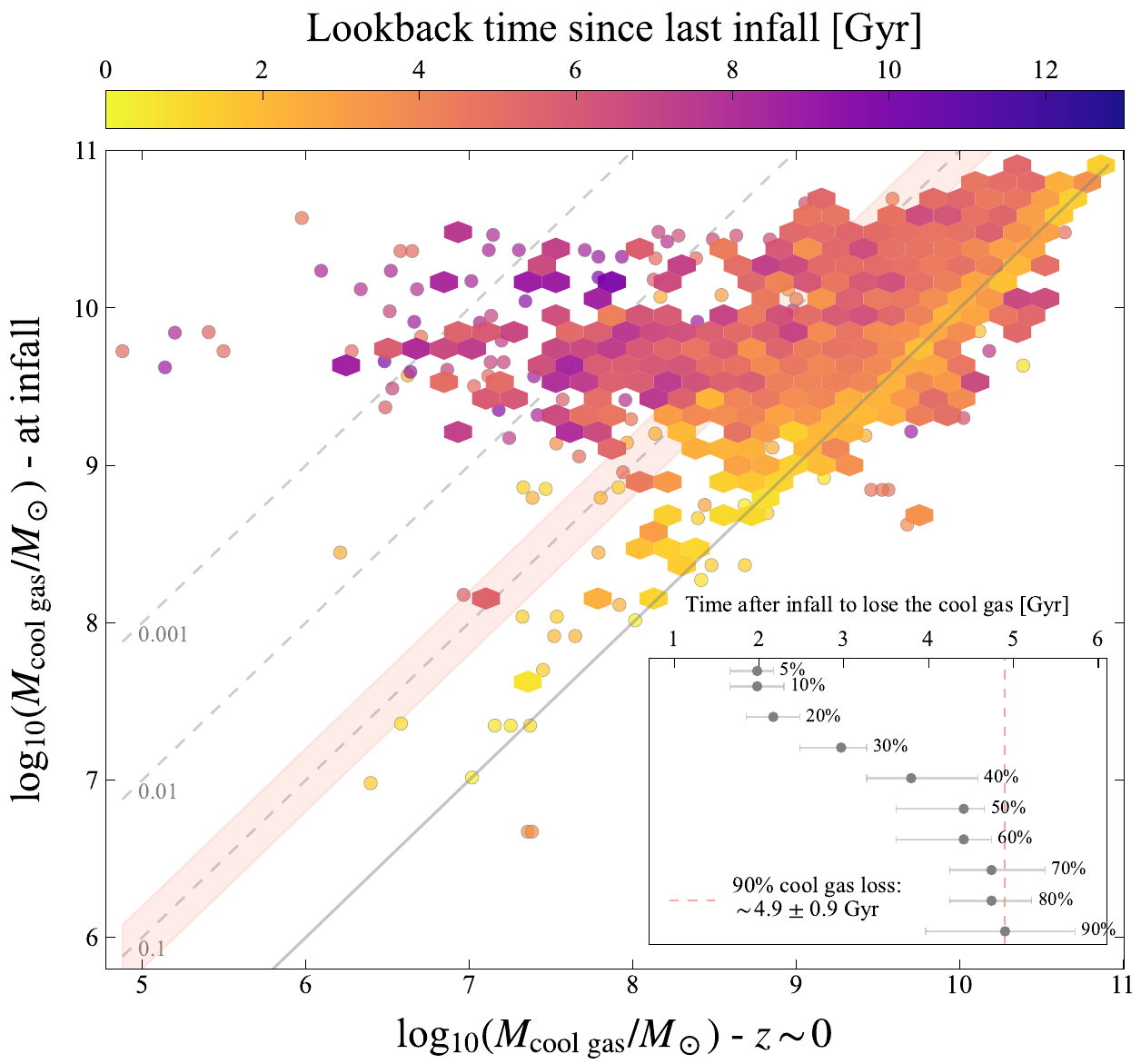}
    \caption{Cool-gas mass of satellites at infall into their present-day host halo versus their current cool-gas mass, color-coded by lookback time since last infall. The solid gray line is the 1:1 relation; dashed lines mark retained cool-gas fractions of $0.1$, $0.01$, and $0.001$ relative to infall. The pink band traces the running median for $t_{90}$. 
    As in Fig.\,\ref{fig:MainPlot}, the inset gives the median time elapsed after infall for satellites to lose successive fractions ($5\%$--$90\%$) of their cool gas, with the red dashed line marking the $90\%$-loss time, $\sim\!4.9^{+0.9}_{-0.9}$\,Gyr.
    Error bars are $3\sigma$ bootstrap uncertainties on the median.}
    \label{fig:coolgas}
\end{figure}

The inset plot in Fig.\,\ref{fig:coolgas}, just like in Figs.\,\ref{fig:MainPlot}\,and\,\ref{fig:Mstar_dependence}, is built directly from these interpolated tracks. For each satellite we define $t_{f}$ as the time after infall at which its cumulative cool-gas loss first reaches a
fraction $f$ of the infall value; taking the population median of $t_{f}$ for
$f=5,10,\dots,90\%$ gives the curve shown, with $3\sigma$ bootstrap errors on each median.
This is the measurement underlying the timescales reported in $\S$\ref{hot_gas}: the median
$t_{90\%}=4.9^{+0.9}_{-0.9}$\,Gyr, the near-flat segment below $f=0.1~(10\%)$ over the first
$\sim\!2$\,Gyr, and the steep rise between $f=0.4~(40\%)$ and $f=0.9~(90\%)$.

As we show in Fig.\,\ref{fig:coolgas}, the depletion is slow at early times---less than 10\% of the cool gas is lost within the first $\sim$2\,Gyr after infall---which is inconsistent with the rapid ($\lesssim$1\,Gyr) removal expected from cool gas stripping processes (e.g., ram-pressure stripping). 
Most of the decline (from 40\% to 90\%) indeed occurs abruptly, within $\sim$1\,Gyr, but only after $\sim$4\,Gyr have elapsed since infall.

\newpage
\section{The $M_{\textrm{hot~gas}}$--SFR relation of centrals}\label{AppendixD}

\begin{figure}[h!]
    \centering \includegraphics[width=0.48\textwidth]{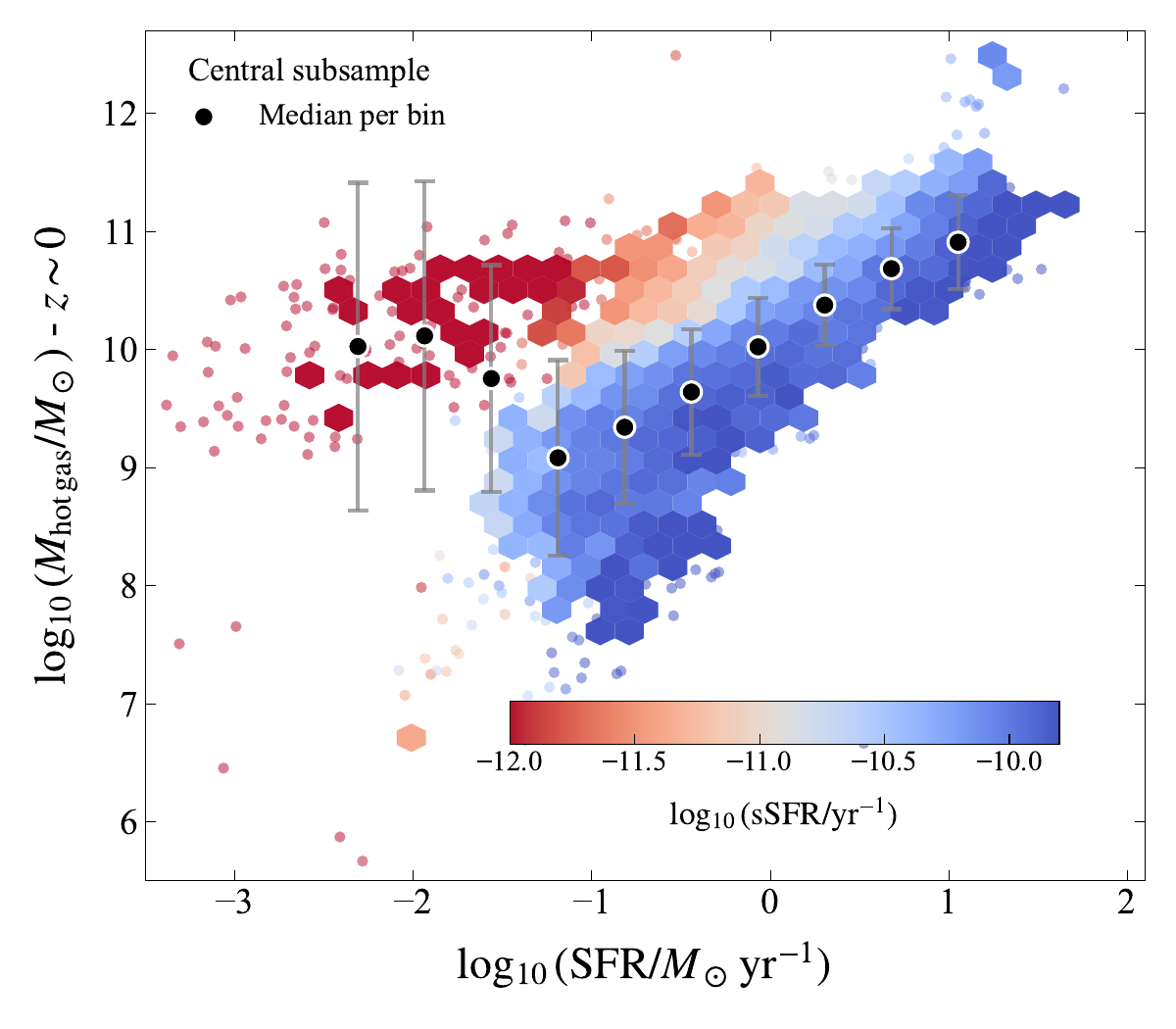}
    \caption{$M_{\rm hot~gas}$--SFR relation for central galaxies at $z$$\sim$0, color-coded by their specific SFR.
    Black circles show the median per SFR bin. Quenched centrals (sSFR$\,\leq10^{-11}$\,yr$^{-1}$) retain large hot reservoirs despite their low SFR, showing that quenching in centrals is not necessarily accompanied by depletion of the hot CGM as verified in satellites.}
    \label{fig:centrals}
\end{figure}

In $\S$\ref{hot_gas} we briefly mention how quenched centrals, unlike satellites, occupy the low-SFR region while simultaneously retaining large hot gas reservoirs.
Figure~\ref{fig:centrals} shows the central subsample in the $M_{\rm hot~gas}$--SFR plane, color-coded by the specific SFR. Two regimes are apparent: star-forming centrals (sSFR$\,>10^{-11}$\,yr$^{-1}$) define a positive, roughly monotonic $M_{\rm hot~gas}$--SFR sequence, analogous to that of the satellites (Fig.\,\ref{fig:MainPlot}), while quenched centrals occupy the low-SFR region but retain large hot reservoirs ($\log_{10}(M_{\rm hot~gas}/M_\odot)\sim10$--$11$). Unlike satellites, therefore, low star formation in centrals is not accompanied by a depletion of the hot CGM.

This points to distinct quenching channels in the two populations: whereas satellite quenching tracks the gradual loss of the hot CGM after infall, centrals retain their hot halo but evidently prevent it from cooling efficiently, most plausibly through maintenance-mode AGN feedback that keeps the gas hot rather than removing it.

\end{appendix}

\end{document}